\documentclass[aps,pra,amssymb,preprint, amsfonts,amsmath,showpacs]{revtex4}
\usepackage{graphicx}
\usepackage{amssymb}
\usepackage{color}
\newcommand{\Tr}{\mathrm{Tr}}
\begin{document}
\title{Heating of trapped ultracold atoms by collapse dynamics}
\author{Franck Lalo\"e}
\email{Franck.laloe@lkb.ens.fr}
\affiliation{Departement de Physique de Laboratoire Kastler Brossel, 
 associ\'{e} de l'ENS, de l'UPMC et du CNRS, 24 rue Lhomond,75005 Paris, France}
\author{William J. Mullin}
\email{mullin@physics.umass.edu}
\affiliation{ Department of Physics, University of Massachusetts, Amherst, MA 01003}
\author{Philip Pearle}
\email{ppearle@hamilton.edu}
\affiliation{Emeritus, Department of Physics, Hamilton College, Clinton, NY  13323}
\pacs{ 03.65.Ta}
\begin{abstract}
 {The Continuous Spontaneous Localization (CSL) theory alters the Schr\"odinger equation. It describes wave function collapse as a dynamical process instead of an ill-defined postulate, thereby providing macroscopic uniqueness and solving the so-called  measurement problem of standard quantum theory.  CSL contains a parameter
 $\lambda$ giving the collapse rate of an isolated nucleon in a superposition of two spatially separated states and, more generally,  characterizing  the collapse time for any physical situation.  CSL is experimentally testable, since it predicts some behavior different from that predicted by standard quantum theory.  One example is the narrowing of wave functions, which results in energy imparted to particles.  Here we consider energy given to trapped  ultra-cold atoms.
Since these are the coldest samples under experimental investigation, it is worth inquiring how they are affected by the CSL heating mechanism.  
We examine the CSL heating of a BEC in contact with its thermal cloud.  
Of course, other mechanisms also provide heat and also particle loss.   From varied data on optically trapped cesium BEC's, we present an energy audit for known heating and loss mechanisms. The result  provides an upper limit on CSL heating and thereby an upper limit on the parameter $\lambda$. We obtain $\lambda\lesssim 1(\pm1)\times 10^{-7}$sec$^{-1}$.}
\end{abstract}
\maketitle

\section{Introduction}

The Continuous Spontaneous Localization (CSL) non-relativistic theory of dynamical collapse \cite{P89},\cite{GPR90} is over two decades old \cite{Reviews}.   It  adds a term to Schr\"odinger's equation, which then includes the collapse of the state vector in its dynamics.  Thus, CSL describes  the occurrence of events, unlike standard quantum theory, which invokes a non-dynamical  `collapse postulate' to account for events.

The added term depends upon a random field $w({\bf x},t)$. One or another  realization of  $w({\bf x},t)$ drives a superposition of states differing in mass density toward one or another of these states (the collapse), thereby accounting for  the world we see around us, which  consists of well-localized mass density.  A second equation, called the `Probability Rule' specifies the probability that nature chooses a particular $w({\bf x},t)$, with the result that the collapse obeys the Born Rule.

 CSL theory contains two parameters, a collapse rate $\lambda$ and a mesoscopic distance $a$.  Suppose a state vector is in a superposition of two states describing a single nucleon at two different locations, with separation $D$. If  $D>>a$, then the superposition collapses at the rate $\approx\lambda/2$.  If  $D<<a$, the collapse rate is $\approx \lambda D^{2}/8a^{2}$. The collapse rate of a state vector describing any number of particles is likewise dependent upon these parameters.  When the superposed states differ macroscopically in mass density, the collapse rate is proportional to $\lambda$ and to the square of the integrated mass density differences, so that it is much larger than $\lambda$.

 If the theory is correct, the  values of the parameters $\lambda$ and $a$ should be determined by experiment.  Provisionally  the parameter values chosen by Ghirardi, Rimini and Weber \cite{GRW} in their instantaneous collapse theory, $\lambda\approx
10^{-16}\hbox {sec}^{-1}$ and $a\approx 10^{-5}\hbox {cm}$, have been adopted.  However,  it should be mentioned that Adler \cite{Adler} has given an argument for $\lambda$ to be as large as $\approx 10^{-8}\hbox {sec}^{-1}$ (and $a\approx 10^{-4}$cm). In Adler's work, as well as in recent articles by Feldmann and Tumulka \cite{Tumulka} and Bassi et. al.\cite{Reviews},
the present and proposed experimental situation has been reviewed.  Currently, the upper limit provided by experiments is $\lambda\lesssim 10^{-9}$sec$^{-1}$\cite{Limit}.

Testing of the theory consists of performing experiments which can yield better limits on the CSL parameters. Either a discrepancy with the predictions of standard quantum theory will appear, validating the theory, or the limits will be such that the theory can no longer be considered viable. For example, this would occur if  $\lambda$ is restricted to be too small, so that the theory does not remove macroscopic superpositions rapidly enough to account for our observations of localized objects.  Such a condition was termed a ``theoretical constraint"  by Collett et. al.\cite{massp} and is referred to by Feldmann and Tumulka\cite{Tumulka} as representing a``philosophically unsatisfactory" condition.

One  situation where CSL makes different predictions than standard quantum theory is the case of bound states.  Here, CSL predicts `spontaneous excitation': particles  will, with a small probability, become excited.  This is because the collapse narrows wave functions, and because a slightly narrowed bound state wave function automatically implies that the state vector becomes the superposition of the initial state plus a small component of  all other accessible states.  Then,  governed by the usual hamiltonian evolution which acts alongside the collapse evolution, the electrons  in atoms or nucleons in nuclei, excited (or ejected) by this collapse, will  radiate (or move away).

    Indeed, experimental limits on these `spontaneous' processes have strongly suggested the mass-density-dependent collapse incorporated in CSL \cite{massp}.  The analysis accompanying these experiments was simplified by utilizing an expansion of the excitation rate in powers of the small parameter (size of bound state/$a)^{2}<<1$.

  In the present paper, we shall apply CSL  in the opposite limit, to ultra-cold atomic gases bound in a magnetic and/or optical  trap. In a sense, these systems are `artificial atoms', where the electrons are replaced by atoms and the nuclear coulomb potential is replaced by a trap with (size of bound state/$a)^{2}>>1$. In this case, the CSL localization effect excites the atoms in the trap.  Various techniques allow experimentalists to give  the trapped atoms
   a temperature as low as a few nanokelvins \cite{CCT-DGO, Stringari-RMP, Pethick-book,Ketterle-picoK}.    One might wonder, since these are the coldest samples studied by physicists, if experimental observations are compatible with the constant heating of the atoms by the CSL process.

The atoms in the trap can assume various forms.     With bosonic samples, experimenters can obtain a Bose-Einstein condensate  (BEC) with a negligible or substantial thermal cloud surrounding the condensate. Experimenters can obtain  a thermal cloud without a condensate (in the case of a normal Fermi gas, of course, the latter is the only possibility).  When a trap is suddenly removed and, after a fixed time  interval, the ensuing atomic density distribution is optically observed,   the number of atoms in the condensate and cloud and  the cloud's velocity distribution, and therefore temperature, may be determined. 

One may envisage experiments designed to detect CSL effects.  CSL predicts the atoms will be heated, resulting in ejection of atoms from a BEC into the cloud. If no other heating process takes place, this results in a  BEC lifetime which is inversely proportional to the CSL collapse rate $\lambda$. 
  
Of course, other well-known mechanisms heat or deplete a Bose-Einstein and its attendant cloud.  Collisions with unavoidable 
 untrapped background gas within the apparatus can heat or eject atoms.  Background photons can produce a similar effect.  There are 3-body collision processes which remove the involved  atoms from the sample.  Atoms in the cloud with energy higher than the trap height can escape the trap. Jitter in the position of the laser beams that form the trap conveys energy to the atoms, as do fluctuations in the laser intensity. 
 
We analyze the contribution of these effects, performing an energy audit on experimental data  kindly supplied by Hanns-Christoph N\"agerl and Manfred Mark for a BEC plus cloud composed of cesium atoms in an optical trap.  This  provides  an upper limit on  $\lambda$. 

The rest of this paper proceeds as follows. Section II  provides a summary of CSL  and a derivation of the evolution equation of the  density matrix for the atoms.  Using this,  in Section III,  the rate equation for the ensemble average of the number of atoms in each bound state  is obtained.
  Section IV explores consequences of the rate equation, applying these results to ideal  situations.  Section V considers real situations.  Section VI applies 
  the considerations of  Section V to  
  particular experiments, thereby obtaining an upper limit on $\lambda$.

\section{CSL Theory}

We first briefly recall the essential features of CSL theory, expressed in terms of the linear solution of the Schr\"odinger equation, rather than the often-employed non-linear Schr\"odinger  stochastic differential equation\cite{P89, GPR90, Reviews}. We shall describe the evolution of state vectors and then utilize that to obtain the evolution of the density matrix, enabling us to study ensemble averaged effects. From the density matrix evolution equation we obtain a rate equation for the state occupation number.  We can then apply the results to the calculation of lifetimes of trapped atomic ultra-cold gases.

\subsection{Summary of CSL theory}

As mentioned in the introduction,
two equations characterize CSL.  The first is a modified Schr\"odinger equation:
\begin{equation}\label{1}
|\psi (t)\rangle_{w}={\cal T}e^{-i\int_{0}^{t}dt'H(t')-\frac{1}{4\lambda}\int_{0}^{t}dt'\int d{\bf x}'[w({\bf x}',t')-2\lambda G({\bf x}')]^{2}}|\psi (0)\rangle.
\end{equation}
\noindent Eq.(\ref{1}) describes collapse toward  the joint eigenstates of mutually commuting operators $G({\bf x})$. ${\cal T}$ is the time-ordering operator, which operates on all operators to its right.  $H$ is the usual hamiltonian.  $G({\bf x})$ is the mass-density \textquotedblleft smeared\textquotedblright\ over a sphere of radius $a$ about the location $\bf x$:
\begin{equation}\label{2}
G({\bf x})\equiv\sum_{n}\frac{m_{n}}{M}\frac{1}{(\pi a^{2})^{3/4}}\int d{\bf z}e^{-\frac{1}{2a^{2}}[{\bf x}-{\bf z}]^{2}}\xi_{n}^{\dagger}({\bf z})\xi_{n}({\bf z}),
\end{equation}
\noindent where $\xi_{n}^{\dagger}({\bf z})$ is the creation operator for a particle of type $n$ at ${\bf z}$, $m_{n}$ is the mass of this particle and $M$ is the mass of a  neutron, and $d{\bf z}\equiv dz_{1}dz_{2}dz_{3}$.  In Eq.(\ref{1}),  $w({\bf x},t)$ is a random field, which at each point of space-time $({\bf x},t)$ can take on any value from $-\infty$ to $\infty$.

The second equation, giving  the probability that nature chooses a particular $w({\bf x},t)$, is the Probability Rule:
\begin{equation}\label{3}
P(w)Dw=_{w}\negthinspace\negthinspace\langle\psi (t)|\psi (t)\rangle_{w}\prod_{{\bf x},t}\frac{dw({\bf x},t)}{\sqrt{2\pi\lambda/dtdv}}
\end{equation}
\noindent where a space-time integral such as appears in Eq.(\ref{1}) is defined as a sum on a discrete lattice of elementary cell volume
$dv$ and time difference $dt$, as the spacing tends to 0.
Since Eq.(\ref{1}) does not describe a unitary evolution of the state vector, the norm of the state vector changes dynamically.  Eq.(\ref{3}) says that state vectors of largest norm are most probable. Of course,  $\int P(w)Dw=1$. This can be seen by
using Eq.(\ref{1}) to insert the state vector norm in Eq.(\ref{3}) and integrating Eq.(\ref{3}) over each $w({\bf x},t)$ from $(-\infty, \infty)$: since  each $w({\bf x},t)$ has a normalized Gaussian distribution, each integral gives 1.

It follows from Eqs.(\ref{1}),(\ref{3}) that the density matrix $\rho(t)$ which describes the ensemble of state vectors evolving under all possible
$w({\bf x},t)'s$  is
\begin{subequations}
\begin{eqnarray}\label{4}
\rho(t) &=&\int P(w)Dw\frac{|\psi (t)\rangle_{ww}\langle\psi(t)|}{_{w}\langle\psi (t)|\psi (t)\rangle_{w}}=\int Dw|\psi (t)\rangle_{ww}\langle\psi (t)|\label{4a}\\
&=&{\cal T}^{+-}e^{-i\int_{0}^{t}dt'(H_{L}(t')-H_{R}(t'))-\frac{\lambda}{2}\int_{0}^{t}dt'\int d{\bf x}'[G_{L}({\bf x}')-G_{R}({\bf x}')]^{2}}\rho(0).\label{4b}
\end{eqnarray}
\end{subequations}
where operators with the subscript $L$ appear to the left of $\rho(0)$ and those with the subscript $R$ appear to the right, and where ${\cal T}^{+-}$  denotes a time ordering operation whereby the $L$ operators are time ordered and the $R$ operators are reverse-time ordered. To derive Eq.(\ref{4b}) from Eq.(\ref{4a}), one can expand both time-ordered exponentials appearing in the right hand side of Eq.(\ref{4a}), perform the  integrals over $w({\bf x},t)$ at each space-time lattice point, and group terms together to obtain the expansion appearing in the exponential of Eq.(\ref{4b}). For this operation,  the integrals over $w$ have   been performed using
\begin{equation}\label{unnumbered}
\int_{-\infty}^{\infty}\frac{dw}{\sqrt{2\pi\lambda/dtdv}}e^{-\frac{1}{4\lambda}dt dv[w-2\lambda G_{L}]^{2}}e^{-\frac{1}{4\lambda}dt dv[w-2\lambda G_{R}]^{2}}=
e^{-\frac{\lambda}{2}dt dv[G_{L}-G_{R}]^{2}}.
\end{equation}

 By taking the time derivative of Eq.(\ref{4}), with use of Eq.(\ref{2}), the density matrix evolution equation  is obtained:
 \begin{eqnarray}\label{5}
\frac{d\rho(t)}{dt}&=&-i[H,\rho(t)]-\frac{\lambda}{2}\sum_{k,n}\frac{m_{k}m_{n}}{M^{2}}
\frac{1}{(\pi a^{2})^{3/2}}\int d{\bf x}\int d{\bf z}\int d{\bf z}'e^{-\frac{1}{2a^{2}}[{\bf x}-{\bf z}]^{2}}e^{-\frac{1}{2a^{2}}[{\bf x}-{\bf z}']^{2}}\cdot\nonumber\\
&&[\xi_{k}^{\dagger}({\bf z})\xi_{k}({\bf z}),[\xi_{n}^{\dagger}({\bf z}')\xi_{n}({\bf z}'),\rho(t)]]\nonumber\\
&=&-i[H,\rho(t)]-\frac{\lambda}{2}\sum_{k,n}\frac{m_{k}m_{n}}{M^{2}}
\int d{\bf z}\int d{\bf z}'e^{-\frac{1}{4a^{2}}[{\bf z}-{\bf z}']^{2}}[\xi_{k}^{\dagger}({\bf z})\xi_{k}({\bf z}),[\xi_{n}^{\dagger}({\bf z}')\xi_{n}({\bf z}'),\rho(t)]]\nonumber\\
&\approx &-i[H,\rho(t)]-\frac{\lambda}{2}
\int d{\bf z}\int d{\bf z}'e^{-\frac{1}{4a^{2}}[{\bf z}-{\bf z}']^{2}}[\xi^{\dagger}({\bf z})\xi({\bf z}),[\xi^{\dagger}({\bf z}')\xi({\bf z}'),\rho(t)]]
\end{eqnarray}
\noindent  In the last step, we have neglected the electron's effect on collapse, which is much smaller than that of the nucleons, neglected the proton-neutron mass difference and also neglected
the distinction between neutrons and protons so $\xi^{\dagger}({\bf z})\xi({\bf z})$ is the number density operator for nucleons.
In the subsequent analysis, we shall only need Eq.(\ref{5}).

\subsection{Density Matrix Evolution Equation for Atoms}

The set of states we shall consider are eigenstates of $H$.  Included in $H$  should be the potential of the externally applied trap and a Hartree-Fock effective potential due to the average influence of all the atoms on one atom (which does not take into account  individual scattering effects). In the calculation that follows, we omit the Hamiltonian term from all expressions and put it back at the end.

We wish to express Eq.(\ref{5}) in terms of the number density operator for atoms (more precisely, its nucleus), instead of the number density operator for individual nucleons as at present.  In the position representation $|{\bf x}\rangle\equiv
|{\bf x}_{1}^{1}, ... {\bf x}_{A}^{1}, ... \hbox{  } {\bf x}_{1}^{N}, ... {\bf x}_{A}^{N}\rangle$ ($A$ is the number of nucleons in each atom's nucleus), Eq.(\ref{5}) becomes
 \begin{equation}\label{6}
\frac{d\langle {\bf x}|\rho(t)|{\bf x}'\rangle}{dt}=
-\frac{\lambda}{2}\sum_{\alpha, \beta=1}^{N}\sum_{i,j=1}^{A}
\Big[ e^{-\frac{1}{4a^{2}}[{\bf x}_{i}^{\alpha}-{\bf x}_{j}^{\beta}]^{2}}+  e^{-\frac{1}{4a^{2}}[{\bf x}_{i}^{'\alpha}-{\bf x}_{j}^{'\beta}]^{2}}
-2e^{-\frac{1}{4a^{2}}[{\bf x}_{i}^{\alpha}-{\bf x}_{j}^{'\beta}]^{2}}\Big]\langle {\bf x}|\rho(t)|{\bf x}'\rangle.
\end{equation}
 
We may label the basis $|{\bf x}\rangle$ in terms of the eigenvalues of the $N$ center of mass operators of the nuclei,
${\bf X}^{\alpha}\equiv A^{-1}\sum_{i=1}^{A}{\bf x}_{i}^{\alpha}$ ($1\leq\alpha\leq N$) and the $(A-1)N$ relative coordinates of the nucleons in each nucleus with respect to its center of mass, ${\bf s}_{i}^{\alpha}\equiv{\bf x}_{i}^{\alpha}-{\bf X}^{\alpha}$ ($1\leq i\leq A-1$:
note, ${\bf s}_{A}^{\alpha}\equiv -\sum_{1=1}^{A-1}{\bf s}_{i}^{\alpha}$ is a dependent variable). Then  Eq.(\ref{6}) becomes
\begin{eqnarray}\label{7}
&&\frac{d\langle {\bf X},{\bf s} |\rho(t)|{\bf X}',{\bf s}'\rangle}{dt}=
-\frac{\lambda}{2}\sum_{\alpha, \beta=1}^{N}\sum_{i,j=1}^{A}
\Big[ e^{-\frac{1}{4a^{2}}[{\bf s}_{i}^{\alpha}-{\bf s}_{j}^{\beta}+{\bf X}^{\alpha}-{\bf X}^{\beta}]^{2}}+
e^{-\frac{1}{4a^{2}}[{\bf s}_{i}^{'\alpha}-{\bf s}_{j}^{'\beta}+{\bf X}^{'\alpha}-{\bf X}^{'\beta}]^{2}}\nonumber\\
&&\qquad\qquad\qquad\qquad\qquad\qquad\qquad\qquad -2e^{-\frac{1}{4a^{2}}[{\bf s}_{i}^{\alpha}-{\bf s}_{j}^{'\beta}+{\bf X}^{\alpha}-{\bf X}^{'\beta}]^{2}}\Big]\langle {\bf X},{\bf s} |\rho(t)|{\bf X}',{\bf s}'\rangle\nonumber\\
&&\approx-\frac{\lambda A^{2}}{2}\sum_{\alpha, \beta=1}^{N}
\Big[ e^{-\frac{1}{4a^{2}}[{\bf X}^{\alpha}-{\bf X}^{\beta}]^{2}}+
e^{-\frac{1}{4a^{2}}[{\bf X}^{'\alpha}-{\bf X}^{'\beta}]^{2}}
 -2e^{-\frac{1}{4a^{2}}[{\bf X}^{\alpha}-{\bf X}^{'\beta}]^{2}}\Big]\langle {\bf X},{\bf s} |\rho(t)|{\bf X}',{\bf s}'\rangle.
\end{eqnarray}
\noindent In the last step, we have used  the fact that the dimensions of the nucleus are very small compared to the CSL parameter $a$ and very small compared to the dimensions of the wave functions we shall consider, so  the ${\bf s}_{i}^{\alpha}$ can be neglected in the exponents.  Finally, we may take the trace of Eq.(\ref{7}) over the relative coordinates, so it becomes an equation for  $\langle {\bf X} |\rho(t)|{\bf X}'\rangle$.  It may then be converted back to an operator equation of the form of Eq.(\ref{5}), expressed in terms of the number density operator for atoms:
   \begin{equation}\label{8}
\frac{d\rho(t)}{dt}=-i[H,\rho(t)]  
-\frac{\lambda A^{2}}{2}
\int d{\bf z}\int d{\bf z}'e^{-\frac{1}{4a^{2}}[{\bf z}-{\bf z}']^{2}}[\zeta^{\dagger}({\bf z})\zeta({\bf z}),[\zeta^{\dagger}({\bf z}')\zeta({\bf z}'),\rho(t)]].
\end{equation}

 Eq.(\ref{8}) is all we use for our calculations.

\section{Rate Equation for Mean Occupation Number}

We consider $N$ atoms bound in a trap; the stationary  states of a
single atom in this trap are described by the wave functions $\varphi
_{i}(\mathbf{x})$ with energy $\omega_{i}$. We include  the trapping potential in $H$ along with the interactions in mean-field approximation,
but ignore spins.   If
$\zeta(\mathbf{x})$ is the field operator, the annihilation operator of the
state $\varphi_{i}({\bf r})$ is:
\begin{equation}\label{1-FL}
a_{i}=\int d{\bf x}\varphi_{i}^{*}({\bf x})\zeta({\bf x})%
\end{equation}
The operator $ N_{i}$\ giving the number of particles in this state and the Hamiltonian $H$ are
\begin{equation}\label{2-FL}
N_{i}\equiv a_{i}^{\dagger}a_{i}=\int d{\bf y}\int d{\bf y}'\varphi^{*}_{i}(\mathbf{y}')\varphi_{i}(\mathbf{y})\zeta^{\dagger}(\mathbf{y})\zeta
(\mathbf{y}'),\medspace H=\sum_{i}\omega_{i}N_{i}
\end{equation}
The operator $Q_{ij}$ whose off-diagonal elements correspond to correlations 
between states and whose diagonal elements are the number operators is:
\begin{equation}\label{3-FL}%
Q_{ij}=a_{i}^{\dagger}a_{j}=\int d\mathbf{y}\int d\mathbf{y}'~\varphi_{i}%
(\mathbf{y})\varphi_{j}^{\ast}(\mathbf{y}')\zeta^{\dagger}(\mathbf{y}%
)\zeta(\mathbf{y}').
\end{equation}

From Eq. (\ref{8}) we get:
\begin{eqnarray}\label{4-FL}%
&& \frac{d}{dt}\Tr\zeta^{\dagger}({\bf y})\zeta({\bf y}')\rho(t)=-i\Tr[\zeta^{\dagger}({\bf y})\zeta({\bf y}'),H]\rho(t)\nonumber\\
&&\qquad\qquad\qquad-\frac{\lambda A^{2}}{2}\Tr\rho(t)
 \int d{\bf z}\int d{\bf z}'e^{-\frac{1}{4a^{2}}[{\bf z}-{\bf z}']^{2}}[\zeta^{\dagger}({\bf z})\zeta({\bf z}),[\zeta^{\dagger}({\bf z}')\zeta({\bf z}'),\zeta^{\dagger}({\bf y})\zeta({\bf y}')]]\nonumber\\
&&=-i\Tr[\zeta^{\dagger}({\bf y})\zeta({\bf y}'),H]\rho(t)\nonumber\\
&&-\frac{\lambda A^{2}}{2}Tr\rho(t)\int d{\bf z}\int d{\bf z}'e^{-\frac{1}{4a^{2}}[{\bf z}-{\bf z}']^{2}}\bigg[\delta({\bf z}-{\bf z}')[\delta({\bf z}'-{\bf y})\zeta^{\dagger}({\bf z})\zeta({\bf y}')
+\delta({\bf z}'-{\bf y}')\zeta^{\dagger}({\bf y})\zeta({\bf z})]\nonumber\\
&&\qquad\qquad\qquad\qquad\qquad\qquad-\delta({\bf z}-{\bf y}')[\delta({\bf z}'-{\bf y})\zeta^{\dagger}({\bf z}')\zeta({\bf z})+\delta({\bf z}'-{\bf y}')\zeta^{\dagger}({\bf z})\zeta({\bf z}')\bigg]\nonumber\\
&&=-i\Tr[\zeta^{\dagger}({\bf y})\zeta({\bf y}'),H]\rho(t)-\lambda A^{2}\Big[1-e^{-\frac{1}{4a^{2}}[{\bf y}-{\bf y}']^{2}}\Big]Tr\zeta^{\dagger}({\bf y})\zeta({\bf y}')\rho(t)
\end{eqnarray}
We note that, when $\mathbf{y}=\mathbf{y}%
^{\prime}$, the left hand side of this equation gives the time derivative of
the ensemble-averaged particle number density, while the collapse part of the right hand side vanishes. The invariance of this 
density during the purely collapse evolution (i.e., evolution when $H=0$) guarantees that
the Born rule is satisfied during collapse towards density eigenstates.

If we multiply both sides of (\ref{4-FL}) by $\varphi_{i}(\mathbf{y}%
)\varphi_{j}^{\ast}(\mathbf{y}')$ and integrate over $d{\bf y}d{\bf y}'$ as in (\ref{3-FL}),
we obtain:%
\begin{eqnarray}
\frac{d}{dt}\langle Q_{ij}\rangle&=&  i(\omega_{i}-\omega_{j})\langle Q_{ij}\rangle\nonumber\\       
&-&\lambda
A^{2}\int d\mathbf{y}\int d\mathbf{y}'~\varphi_{i}(\mathbf{y})\varphi
_{j}^{\ast}(\mathbf{y}')\left[  1-e^{-\left(  \mathbf{y}-\mathbf{y}'\right)
^{2}/4a^{2}}\right]  \Tr\zeta^{\dagger}({\bf y})\zeta({\bf y}')\rho(t)\label{5-FL}%
\end{eqnarray}
\noindent where the notation $\Tr Q_{ij}\rho(t)\equiv\langle Q_{ij}\rangle$ and $\Tr N_{i}\rho(t)\equiv\langle N_{i}\rangle$ is employed.
 Using the inversion of 
Eq.(\ref{1-FL}) and its hermitian conjugate,   $\zeta({\bf y}')=\sum_{k'}a_{k'}\varphi_{k'}({\bf y}')$ and $\zeta^{\dagger}({\bf y})=
\sum_{k}a_{k}^{\dagger}\varphi_{k}^{*}({\bf y})$,
we may write $\zeta^{\dagger}({\bf y})\zeta({\bf y}')=\sum_{kk'}Q_{kk'}\varphi_{k}^{*}({\bf y})\varphi_{k'}({\bf y}')$. Inserting this into Eq.(\ref{5-FL}) results in
\begin{equation}
\frac{d}{dt}\langle Q_{ij}\rangle=  i(\omega_{i}-\omega_{j})\langle Q_{ij}\rangle    
-\lambda A^{2}\sum_{kk'}\gamma_{ij}^{kk'}\langle Q_{kk'}\rangle
\label{6-FL}%
\end{equation}
\noindent where the coefficients $\gamma_{ij}^{kk^{\prime}}$ are defined by:%
\begin{equation}
\gamma_{ij}^{kk^{\prime}}=\int d\mathbf{y}\int d\mathbf{y}'\left[  1-e^{-\left(
\mathbf{y}-\mathbf{y}^{\prime}\right)  ^{2}/4a^{2}}\right]  
\varphi_{i}(\mathbf{y})\varphi_{j}^{*}(\mathbf{y}')
\varphi_{k}^{*}(\mathbf{y})\varphi_{k^{\prime}}(\mathbf{y}').%
 \label{9-FL}%
\end{equation}
\noindent This provides the rate equations describing the dynamics of the system.  

 In particular,  we are interested in 
 the time evolution of the mean number of particles in the ith state, $\langle Q_{ii}\rangle=\langle N_{i}\rangle$.  But, 
 we see from Eq.(\ref{6-FL}) that this diagonal element is coupled to the off-diagonal elements.

However, these equations can be simplified if we assume that $\lambda A^{2}$ is much
smaller than all frequency differences $\left(  \omega_{i}-\omega_{j}\right)
$, which holds for the applications discussed in this paper. When $\lambda A^{2}$ is small, the populations (terms $i=j$) evolve slowly, only
under the effect of the CSL process, while the off-diagonal terms tend to
oscillate at high frequencies $\left(  \omega_{i}-\omega_{j}\right)  $.
Because this oscillation is too fast, the time integrated effect of these
off-diagonal terms on the populations then averages out to almost perfectly
zero, and can therefore be ignored (secular approximation).\ We then obtain the following evolution equations for the populations:%
\begin{equation}
\frac{d}{dt}\langle N_{i}\rangle\simeq-\lambda
A^{2}\sum_{k}\gamma_{i}^{k}\langle N_{k}\rangle\label{19-FL}%
\end{equation}
with:%
\begin{equation}
\gamma_{i}^{k}=\int d\mathbf{y}\int d\mathbf{y}'\left[  1-e^{-\left(
\mathbf{y}-\mathbf{y}^{\prime}\right)  ^{2}/4a^{2}}\right]  
\varphi_{i}(\mathbf{y})\varphi_{i}^{*}(\mathbf{y}')
\varphi_{k}^{*}(\mathbf{y})\varphi_{k}(\mathbf{y}').%
 \label{20-FL}%
\end{equation}
These equations provide a closed system for the evolution of the populations, 
in the form of coupled linear rate equations. 
The solution to the rate equations is discussed in Appendices B and C.  The rest of this paper deals with the consequences of the  rate equations unmodified, or modified (by the addition of thermalizing collisions or an external influence).  

As an example of the consequences of the unmodified rate equations, consider an initial BEC of $N$ atoms in a three-dimensional spherically symmetric harmonic trap.  
The occupation numbers  of the cloud  states as a function of time are given by Eq.(\ref{CBEC1b}) of Appendix C, and are plotted in Fig. \ref{figFittedData}. 

\begin{figure}[h]
\includegraphics[width=3.5in]{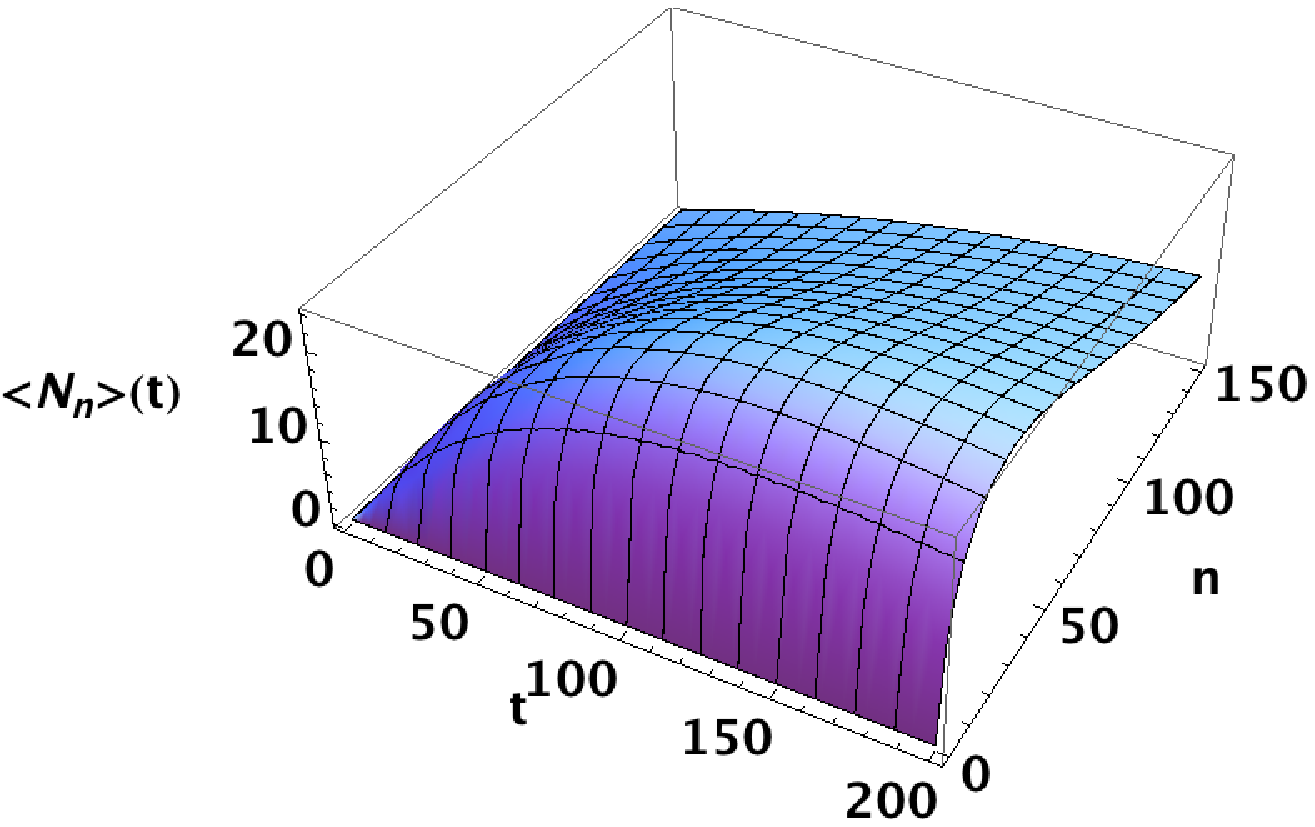}$\quad\quad$\includegraphics[width=2.5in]{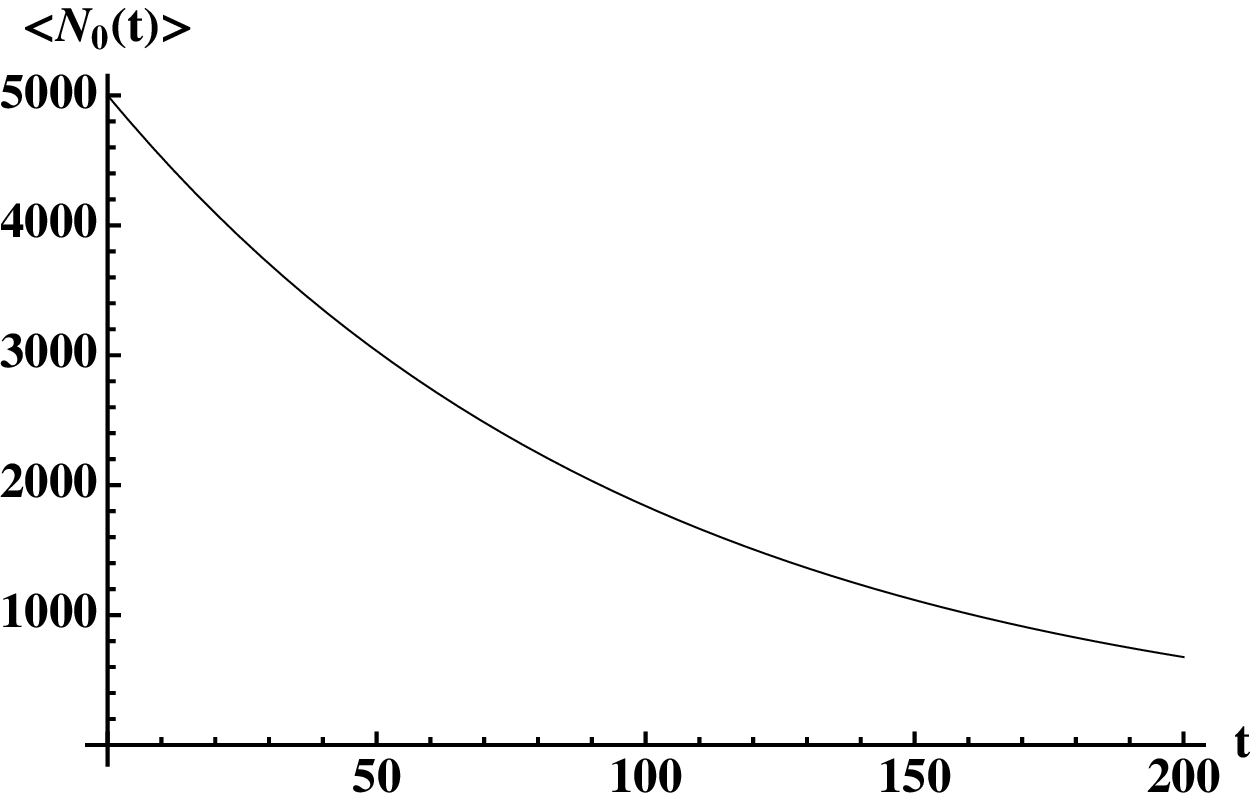}
(Color online)\caption{(Left)  (Color online) Starting from an initial pure BEC of $N=5000$ bosons, the
plot shows the mean occupation number $\langle N_{n}\rangle(t)$ of
state number $n$ ($n\neq0)$ as a function of time $t$ (arbitrary units) due to CSL heating. We
take $\lambda A^{2}=0.01$ and $\alpha=a/\sigma=0.1$, where $\sigma$
is the harmonic oscillator characteristic length $\sqrt{\hbar/m\omega}$.
The energy $\varepsilon$ is related to occupation number via $\varepsilon=\hbar\omega n$.
(Right) The exponential loss $\langle N_{0}\rangle(t)=N(0)e^{-\lambda A^{2} t}$ of the ground state
with CSL heating. }

\label{figFittedData}
\end{figure}

\section{Some Consequences of Rate Equations}

We shall first discuss general properties of the rate equations.  Then, we shall consider some detailed consequences. 

When Eq.(\ref{19-FL}) is summed over $i$, the right hand side vanishes, reflecting the constant value of the total number of atoms $\equiv N$ in all states.  This is because 
$\sum_{i}\gamma_{i}^{k}=0$, which  follows from $\sum_{i}\varphi_{i}(\mathbf{y})\varphi_{i}^{*}(\mathbf{y}')=\delta({\bf y}-{\bf y}')$.    

If $i=k$, the integrand in the definition of $\gamma_{i}^{i}$ is positive,
which means that the CSL\ self-coupling coefficient of any population is
always negative, corresponding to its decay.

Using  the Fourier transform of the exponential in Eq.(\ref{20-FL}), $\gamma_{i}^{k}$ may be written as
\begin{equation} \label{Rate1}
\gamma_{i}^{k}=\delta_{ik}-\frac{a^{3}}{\pi^{3/2}}\int d{\bf q}e^{-{\bf q}^{2}a^{2}}\Big|\int d{\bf y}e^{i{\bf q}\cdot{\bf y}}\varphi_{i}(\mathbf{y})\varphi_{k}^{*}(\mathbf{y})\Big|^{2}.
\end{equation}
\noindent  Thus, if $i\neq k$, $\gamma_{i}^{k}$ is negative. 
By replacing $e^{-{\bf q}^{2}a^{2}}$ by 1 and performing the ${\bf k}$ integral, we obtain an  upper bound  
on this term:
\[
(2a\sqrt{\pi})^{3}\int d{\bf y}|\varphi_{i}(\mathbf{y})|^{2}|\varphi_{k}(\mathbf{y})|^{2}=C(a/\sigma)^{3}
\]
\noindent where we take the scale of the wave function to be $\sigma$.  Since it is assumed that $\sigma>>a$,  according to Eq.(\ref{19-FL}), all initial state populations decay at a rate slightly less than $\lambda A^{2}$ but, in each interval $dt$, the $i$th state repopulates the rest (including itself) with a positive fraction $\sim \lambda A^{2}dt\langle N_{i}\rangle$.  

It is an exact property of CSL that, regardless of the potential in the Hamiltonian, the mean energy of any system increases linearly with time.  As shown in Appendix A:
\begin{equation}\label{Rate2.5}
\frac{d}{dt}\overline E=\lambda A^{2}N\frac{3\hbar^{2}}{4ma^{2}}.
\end{equation}
 
 So,  the initial bound state distribution of the $N$ atoms gets excited at the rate  $\lambda A^{2}$ with an average energy increase per atom 
 $\sim \hbar^{2} / ma^{2}\equiv k_{B}T_{CSL}$, where $k_{B}$ is Boltzmann's constant.   If $a= 10^{-5}$cm, for the two atoms $^{87}$Rb and $^{133}$Cs which are frequently used to form condensates, $ k_{B}T_{CSL}$ has the respective values $\approx 550$nK and $\approx 360$nK. 
 
 \subsection{Bose-Einstein Condensate Without Thermal Cloud}
 
Suppose one starts with all $N$ particles in the Bose-Einstein condensate ground state $\varphi_{1}(\mathbf{y})$, and sweeps away particles which are ejected (or achieves the same effect by making a very shallow trap with $T<<T_{CSL}$). Then, the rate equation (\ref{19-FL}) for  $\langle N_{1}\rangle$ is to be modified so that the excited states do not  inject atoms back into the BEC, and only the term $\gamma_{1}^{1}$  contributes : 
 \begin{equation}\label{Rate3}
\frac{d}{dt}\langle N_{1}\rangle=     -\lambda
A^{2}\Big[1-\int d\mathbf{y}\int d\mathbf{y}'e^{-\left(
\mathbf{y}-\mathbf{y}^{\prime}\right)  ^{2}/4a^{2}} 
|\varphi_{1}(\mathbf{y})|^{2}|\varphi_{1}(\mathbf{y}')|^{2}\Big]\langle N_{1}\rangle.
\end{equation}
\noindent Since the ground state wave function scarcely changes over the distance $a$, it is a good approximation to replace the exponential in Eq.(\ref{Rate3})
by a delta function, obtaining:
 \begin{equation}\label{Rate4}
\frac{d}{dt}\langle N_{1}\rangle\approx    -\lambda
A^{2}\Big[1-a^{3}[4\pi ]^{3/2}\int d\mathbf{y}
|\varphi_{1}(\mathbf{y})|^{4}\Big]\langle N_{1}\rangle=  -\lambda
A^{2}\Big[1-[4\pi ]^{3/2}C\Big(\frac{a}{\sigma}\Big)^{3}\Big]\langle N_{1}\rangle
\end{equation}
\noindent where $C$ is of order 1: for a box of side length $\sigma$, $C=(3/2)^{3}$ and for a harmonic oscillator potential with 
$\sigma\equiv(m\omega)^{-1/2}$, $C=(2\pi)^{-3/2}$.  

Thus, to an excellent approximation, the lifetime of the BEC due to the CSL process alone is 
$\tau_{CSL}\approx 1/(\lambda A^{2})$.  If this experiment were to be performed,  with a measured lifetime $\tau_{exp}$ resulting from the CSL process together with all other processes that limit the lifetime, since $\tau_{CSL}\geq\tau_{exp}$, that would place a limit  
 \begin{equation}\label{BECLimit}
\lambda<1/(\tau_{exp}A^{2}). 
\end{equation}
For example, if the experiment was done with $^{133}$Cs and the measured lifetime was $\tau_{exp}=10$s, the limit would be 
$\lambda<6\times 10^{-6}$s$^{-1}$.

The standard external optical or magneto-optical potential is harmonic, although recently a cylindrical box potential has been used\cite{Hadz}. However, the \textit{effective} potential is neither a box nor a harmonic oscillator potential. 
Usually a condensate is so dense that it creates a potential that dominates the trap's harmonic oscillator potential near the center of the trap.  The wave function in this case is given by  solving the  time-independent Schr\"odinger equation with both potentials, which is called the Gross-Pitaevskii equation. To a good approximation (called the Thomas-Fermi approximation),  the squared wave function for a BEC atom is\cite{Baym}
 \[
\phi_{0}^{2}({\bf y})=\frac{1}{\sigma^{3}}\Big[\Big(\frac{15}{8\pi}\Big)^{2/5}-\frac{{\bf y}^{2}}{\sigma^{2}}\Big]
\]
\noindent where positive, and 0 for larger ${\bf y}^{2}$:  $\sigma$ is a large  multiple of the harmonic oscillator width.  Putting this into (\ref{Rate4}), one obtains 
$C\approx 1.7$.

\subsection{Thermal  Cloud Without Bose-Einsein Condensate}

The atomic energy scale in a  trap potential  characterized by length $\sigma$ is $\hbar^{2}/m\sigma^{2}$ while the CSL energy scale is $\hbar^{2}/ma^{2}$.  Since  $\alpha\equiv a/\sigma<<1$, the CSL excitation covers many atomic states, so a sum over states can be well-approximated by an integral. For a sparse thermal cloud, the   effective potential is the harmonic trap potential. If the thermal cloud is dense, and the atomic force is repulsive (which we shall assume, e.g., for $^{87}$Rb and 
$^{133}$Cs, the s-wave scattering length is positive), the effective potential tends to be flattened at the bottom, suggesting the utility of a calculation based upon a box potential.

For either the spherically symmetric  harmonic oscillator potential   with $\hbar\omega=\hbar^{2}/m\sigma^{2}$, or for the box with side length $\sigma$, we show in the appendices that both yield the same rate equations (\ref{A9}) or (\ref{B10}), for  $\langle N_{\epsilon}\rangle$, the number of states per unit energy range,
\begin{equation}\label{Univ}
\frac{d}{dt}\langle N_{\epsilon}\rangle=-\lambda A^{2}\langle N_{\epsilon}\rangle+\lambda A^{2}\frac{1}{\sqrt{\pi k_{B}T_{CSL}/2}}\int_{0}^{\infty}d\sqrt{\epsilon'}
\Big[e^{-\frac{2}{k_{B}T_{CSL}/2}(\sqrt{\epsilon}-\sqrt{\epsilon'})^{2}}-e^{-\frac{2}{k_{B}T_{CSL}}(\sqrt{\epsilon}+\sqrt{\epsilon'})^{2}}\Big]\langle N_{\epsilon'}\rangle.
\end{equation}
\noindent with solution Eq.(\ref{A9})
\begin{eqnarray}\label{SERSOL}
 \langle N_{\epsilon}\rangle(t)
&=& \langle N_{\epsilon}\rangle(0)e^{-\lambda A^{2}t}\nonumber\\
&+&\frac{1}{\sqrt{\pi k_{B}T_{CSL}/2}}\int_{0}^{\infty}d\sqrt{\epsilon'}\langle N_{\epsilon'}\rangle(0)\sum_{s=1}^{\infty}e^{-\lambda A^{2}t}\frac{(\lambda A^{2}t)^{s}}{s!\sqrt{s}}
\Big[e^{-\frac{2}{sk_{B}T_{CSL}}(\sqrt{\epsilon}-\sqrt{\epsilon'})^{2}}-e^{-\frac{2}{sk_{B}T_{CSL}}(\sqrt{\epsilon}+\sqrt{\epsilon'})^{2}}\Big].\nonumber\\\label{C3c}
\end{eqnarray}
\noindent As expanded upon in Appendix B, this rate equation and solution are expected to be  good approximations for any trap where the potential energy is negligibly small compared to the kinetic energy $k_{B}T_{CSL}$ for most of the spatial range of the wave function, which is the case for traps used in BEC experiments.

\begin{figure}[h]
\centering \includegraphics[width=4in]{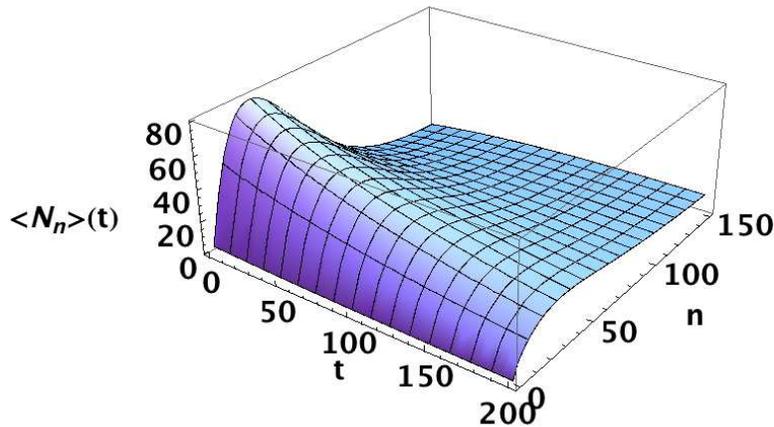}

\caption{ (Color online) Starting from an initial thermal cloud of $N=5000$ bosons at  temperature $85$nK in a three-dimensional spherically symmetric harmonic trap, the plot shows the mean occupation number $\langle N_{n}\rangle(t)$ of  state number $n$ as a function of 
time $t$ (arbitrary units) due to CSL heating. In this example, the transition temperature is $T_{c}=80.4$nK so there is no  BEC, just a cloud. We take $\lambda A^{2}=0.01$, $T_{CSL}=500$ nK and $\alpha=\sqrt{\hbar\omega/k_{B}T_{CSL}}=.1$, so  $\hbar\omega/k_{B}=5$nK. }

\label{fig2}
\end{figure}

To illustrate  Eq.(\ref{SERSOL}),  Fig.\ref{fig2} displays the decay of an initial thermal cloud 
distribution into a time-dependent cloud generated by CSL heating. 
The energy $\varepsilon$ has been converted to
occupation number via $\varepsilon=\hbar\omega n$. 

This shows how the higher energy states are populated by the CSL heating mechanism as time increases.  Interaction between the atoms has been disregarded here: of course, when it is included, the cloud continuously thermalizes and so may be characterized by an increasing temperature.

\subsection{Bose-Einstein Condensate and Thermal Cloud I}

A common experimental situation is a BEC in thermal equilibrium with its surrounding cloud.  We shall first consider the special case of noninteracting atoms in an 
anisotropic harmonic oscillator potential of infinite height. For completeness, its well known statistics are reviewed before 
use in our specific application.  

The density of states is $\epsilon^{2}/2\overline\omega^{3}$, where  $\epsilon$ is the system energy and $\overline\omega\equiv 2\pi(f_{1}f_{2}f_{3})^{1/3}$ (the oscillator's three frequencies are $f_{1},f_{2},f_{3}$).  The BEC occupies the ground state whose energy we take to be 0.  
 If there are N atoms,  Bose-Einstein statistics implies,  below the critical temperature $T_{c}$, that the cloud contains $N_{cloud}<N$ atoms given by 
\begin{equation}\label{BECC1}
N_{cloud}=\frac{1}{2\overline\omega^{3}}\int_{0}^{\infty}d\epsilon\frac{\epsilon^{2}}{e^{\frac{\epsilon}{k_{B}T}}-1}=
\frac{1}{2}\Big(\frac{k_{B}T}{\overline\omega}\Big)^{3}\int_{0}^{\infty}dy\frac{y^{2}}{e^{y}-1}=\Big(\frac{k_{B}T}{\overline\omega}\Big)^{3}\zeta(3)
\end{equation}
  \noindent  where $\zeta$ is the Riemann zeta function, with
   $\zeta(3)=1.20...$ . This calculation utilizes the approximation $\hbar\overline \omega<<k_{B}T$   in replacing the sum over discrete states by an integral over the  density of states.

  At the critical temperature $T_{c}$ below which the BEC forms, all the atoms are in the cloud so,  from Eq.(\ref{BECC1}), we have
  \begin{equation}\label{BECC2}
N=\Big(\frac{k_{B}T_{c}}{\overline\omega}\Big)^{3}\zeta(3). 
\end{equation}
\noindent Therefore, for $T< T_{c}$, from Eqs.(\ref{BECC1}),(\ref{BECC2}) it follows that 
   \begin{equation}\label{BECC3}
N_{cloud}=N\Big(\frac{T}{T_{c}}\Big)^{3}, \qquad N_{BEC}=N\Big[1-\Big(\frac{T}{T_{c}}\Big)^{3}\Big].
\end{equation}
The condensate fraction is $f\equiv N_{BEC}/N$.
  
Similarly, Bose-Einstein statistics gives the cloud energy  and specific heat/atom:
\begin{eqnarray}\label{BECC4}
U_{cloud}&=&3k_{B}T\Big(\frac{k_{B}T}{\overline\omega}\Big)^{3}\zeta(4)=N3k_{B}T_{c}\Big(\frac{T}{T_{c}}\Big)^{4}\frac{\zeta(4)}{\zeta(3)}\nonumber\\
C&\equiv&\frac{1}{N}\Big(\frac{\partial U(T,N)}{\partial T}\Big)_{N}=\frac{1}{N}\Big(\frac{\partial U_{cloud}(T,N)}{\partial T}\Big)_{N}=12k_{B}\Big(\frac{T}{T_{c}}\Big)^{3}\frac{\zeta(4)}{\zeta(3)}
\end{eqnarray}
\noindent where $\zeta(4)=1.08...$ . Note that $U_{cloud}(T,N)$ depends only on $T$ because $T_{c}^{3} \sim N$.

Using conservation of energy, if only the CSL heating mechanism operates, the rate of increase of the CSL energy (\ref{Rate2.5})  equals the rate of increase of $U$:
\begin{equation}\label{BECC5}
	N\lambda A^{2}\frac{3}{4}k_{B}T_{CSL}=\frac{d}{dt}U_{cloud}=NC\frac{dT}{dt}=NC\frac{dT}{df}\frac{df}{dt}=-\frac{NCf}{\tau_{f}df/dT}. 
\end{equation}
\noindent  Eq.(\ref{BECC5}) has been expressed in terms of the lifetime $\tau_{f}$ of the condensate fraction:
\begin{equation}\label{BECC6}
\frac{1}{\tau_{f}}\equiv-\frac{1}{f}\frac{df}{dt }
\end{equation}
\noindent because this is a readily measurable quantity.  Inserting into (\ref{BECC5}) expression (\ref{BECC4}) for $C$ and (\ref{BECC3}) for $F$, and solving for $\lambda$, we obtain
\begin{equation}\label{BECC7}
\lambda = \frac{1}{A^{2}\tau_{f}}\frac{T}{T_{CSL}}\Bigg[1-\Big(\frac{T}{T_{c}}\Big)^{3}\Bigg]\frac{16\zeta(4)}{3\zeta(3)}.
\end{equation}
\noindent Therefore,  were CSL to provide the only heating effect, one could measure $\lambda $ by measuring the BEC fraction lifetime and the temperature.

\subsection{Bose-Einstein Condensate and Thermal Cloud II}

Atoms do interact, so the potential felt by each atoms is not just the harmonic oscillator potential of the trap.  Reference \cite{Stringari-RMP} considers an interacting Bose gas in the finite-temperature Hartree-Fock scheme with the Thomas-Fermi approximation for the condensate.  The equations in the previous section are modified in this case as follows. The chemical potential $\mu$ is no longer 0:
 
\begin{equation}\label{HF1}
\mu=k_{B}T_{c}\eta(1-s^{3})^{2/5}
\end{equation}
where 
\begin{equation}\label{HF2}
\eta=\frac{\hbar\omega}{2k_{B}T_{c}}\left(\frac{15Na_{s}}{a_{HO}}\right)^{2/5}.
\end{equation}
\noindent Here,  $a_{HO}=\sqrt{\hbar/m\omega}$, $a_{s}$ is the atom-atom s-wave scattering length and $s\equiv T/T_{c}$: note that 
the  critical temperature (at constant number of atoms)  does not keep the same value as for an ideal gas, because of the change of density 
at the center of the trap induced by the interaction, but these equations use the expression (\ref{BECC2}) for the ideal gas $T_{c}$.

 The  condensate  fraction is
\begin{equation}\label{HF3}
f=1-s^{3}-\frac{\zeta(2)}{\zeta(3)}\eta s^{2}(1-s^{3})^{2/5}
\end{equation}
and the energy is 
\begin{equation}\label{HF4}
U=Nk_{B}T_{c}\Bigg\{\frac{3\zeta(4)}{\zeta(3)}s^{4}+\frac{1}{7}\eta(1-s^{3})^{2/5}(5+16s^{3})\Bigg\}.
\end{equation}

From Eqs.(\ref{HF3}),(\ref{HF4}) we can evaluate the following quantities:
\begin{eqnarray}\label{HF5}
C & = &\frac{1}{N}\Big(\frac{\partial U}{\partial T}\Big)_{N}=k_{B}\Bigg\{ \frac{12s^{3}\zeta(4)}{\zeta(3)}+\frac{6s^{2}\eta}{\left(1-s^{3}\right)^{3/5}}\left(1-\frac{56s^{3}}{35}\right)\Bigg\}\\
\mu_{N}&\equiv&\Big(\frac{\partial U}{\partial N}\Big)_{T}=k_{B}T_{c} \frac{\eta}{5(1-s^{3})^{3/5}}(5+s^{3})\\
f_{T} & \equiv & \Big(\frac{\partial f}{\partial T}\Big)_{N}=           -\frac{3s^{3}}{T}-\frac{2\eta\zeta(2)s^{2}}{5T\zeta(3)\left(1-s^{3}\right)^{3/5}}\left(5-8s^{3}\right)\\
f_{N} & \equiv &N\Big(\frac{\partial f}{\partial N}\Big)_{T}= s^{3}+\frac{\eta\zeta(2)s^{2}}{15\zeta(3)\left(1-s^{3}\right)^{3/5}}\left(9-15s^{3}\right)
\end{eqnarray}
In all cases, these expressions differ from those in the previous section by a term $\sim\eta$. 

\subsection{Bose-Einstein Condensate and Thermal Cloud III}
We now wish to obtain an expression for $\lambda$ similar to (\ref{BECC7}) for the case of interacting bosons and also now allow for 
the loss of atoms (change of $N$) as occurs in actual experiments. In so doing, the result shall be expressed in terms of practically measurable quantities. 

The rate of increase of $U$ is given by
\begin{equation}\label{Pr1}
\frac{1}{N}\frac{d}{dt}U=C\frac{dT}{dt}+\mu_{N}\frac{1}{N}\frac{dN}{dt}
\end{equation}
We shall evaluate this equation at $t=0$. Graphs of $N(t)$ and $f(t)$ may be experimentally obtained, from which the initial slopes may be extracted.  These are defined as

\noindent 
\begin{eqnarray}\label{eq:tauf}
\frac{1}{\tau_{N}} & \equiv& -\left.\left(\frac{1}{N}\frac{dN}{dt}\right)\right|_{t=0}\label{eq:tauN}\\
\frac{1}{\tau_{f}} & = & =-\left.\left(\frac{1}{f}\frac{df}{dt}\right)\right|_{t=0}
\end{eqnarray}
We do not assume the time dependence is strictly exponential for either
$N$ or $f.$ We have
\begin{equation}\label{eq:dF}
\left.\frac{df}{dt}\right|_{t=0}=-\frac{f(0)}{\tau_{f}}=f_{T}\left.\frac{dT}{dt}\right|_{t=0}+f_{N}\frac{1}{N}\left.\frac{dN}{dt}\right|_{t=0}
\end{equation}
\noindent With all quantities evaluated at $t=0$, Eq. (\ref{eq:dF}) implies that
\begin{equation}\label{eq:dT}
\frac{dT}{dt}=\frac{1}{f_{T}}\left[\frac{f_{N}}{\tau_{N}}-\frac{f}{\tau_{f}}\right]
\end{equation}
Substituting this into Eq. (\ref{Pr1}) gives
\begin{equation}\label{eq:LHS}
\frac{1}{N}\frac{dU}{dt}=\frac{C}{f_{T}}\left[\frac{f_{N}}{\tau_{N}}-\frac{f}{\tau_{f}}\right]-\frac{\mu_{N}}{\tau_{N}}
\end{equation}

We emphasize that Eq.(\ref{eq:LHS}) is expressed in terms of experimental quantities and Hartree-Fock calculated quantities given in the previous section.  It holds 
 regardless of the specific mechanisms that heat or cool the atoms, or that remove atoms from the trap. In the following sections 
 we shall calculate the contributions of various heating mechanisms in addition to that of CSL, and equate their energy change per particle to (\ref{eq:LHS}). 
 
As a simple application, were there no other heating source other than CSL, as in Eq.(\ref{BECC5}), 
 the rate of increase of the CSL energy  (\ref{Rate2.5}) equals the
rate of increase of $U$:
\begin{equation}
N\lambda A^{2}\frac{3}{4}k_{B}T_{CSL}=\frac{dU}{dt}.  
\end{equation}
It follows that the equivalent of Eq.(\ref{BECC7})  is
\begin{equation}\label{La1}
\lambda=\frac{4}{3}\frac{1}{A^{2}k_{B}T_{CSL}}\left\{ \frac{C}{f_{T}}\left[\frac{f_{N}}{\tau_{N}}-\frac{f}{\tau_{f}}\right]-\frac{\mu_{N}}{\tau_{N}}\right\}, 
\end{equation}
where the quantities in this equation are to be obtained from the Hartree-Fock expressions of the previous section. Of course, 
 Eq.(\ref{La1}) is identical to Eq.(\ref{BECC7}) when $\eta=0$. 

\section{CSL Heating with External Heating and Loss of Atoms}

Naturally, in any actual experiment, in addition to heating from CSL, there will be other heating and cooling
sources,  and we consider this most general situation here.
The heating processes we consider (the first two we found to be most significant) are as follows. 

1)  The rate of atoms in the cloud leaving the trap, with energy greater than the trap
barrier height $\epsilon_{w}$,   is $1/\tau_{cool}$.
These processes are the source of evaporative cooling. The
theory of Luiten et al \cite{LRW} evaluates $\tau_{cool}$
and the cooling power, which we write as $(dU/dt)_{cool}$ (a negative
quantity). 

2)  Three-body recombination (TBR) occurs when three atoms in the BEC inelastically collide, 
two forming a dimer, but all typically departing the trap when the barrier is low.  BEC experiments usually try to minimize 
TBR losses.  However, in the experimental data we  examine, TBR
is \emph{not} negligible. Indeed,  it is apparently the primary determinant of $\tau_{N}$ (dominating the particle number loss in 1) above), 
and contributes heating to the gas given by $(dU/dt)_{TBR}.$ We
 estimate the value of this\cite{Weber}. We find that, in order to fit the data,  the TBR decay curve 
must be accompanied by a exponential tail at long times.  We assume this tail is due to the evaporative cooling described in 1) above, and this gives us a value for  $\tau_{cool}$.

3) Foreign atoms in the vacuum chamber, often mostly hydrogen, at essentially room temperature, occasionally collide with the atoms in the BEC or thermal cloud.  With high probability, once struck, a Cs atom leaves the trap with no further collisions. We assume that the rate of atoms lost per atom, denoted $\tau_{1}^{-1}$, 
is the same for atoms in the BEC as it is for atoms in the thermal cloud.  We denote the energy lost per atom as $-U_{av}/N\tau_{1}$, where $U_{av}$ is the average energy per atom in the system. $\tau_{1}$ can be estimated from the literature\cite{Bali}.
 
4) Struck atoms which do not escape from the trap distribute their received energy. 

5)  Cs atoms removed from the trap may still occupy the neighborhood 
(the so-called ``Oort cloud" \cite{Cornell}) and collide with trapped atoms. 

6) Mechanical jitter of the laser beam focus which traps the atoms can heat them up.  Intensity fluctuations of the laser beam have a similar effect.\cite{Savard}  

We shall denote by  $R_{in}$  the rate of heating per atom due to CSL and sources 4)-6) (which we estimate as small  but do not bother to provide the estimation here) and any other or unknown sources: the latter 
 is basically what we will find as the residual in an experiment. 

Then, equating the experimentally measurable energy change given by Eq.(\ref{eq:LHS}) to these listed sources of energy  results  in the relation
\begin{equation}\label{CALC}
\frac{1}{N}\frac{dU}{dt}=R_{in}-\frac{1}{\tau_{1}}\frac{U_{av}}{N}+\frac{1}{N}\left.\frac{dU}{dt}\right|_{cool}+\frac{1}{N}\left.\frac{dU}{dt}\right|_{TBR}
\end{equation}
with, of course, $dU/dt|_{cool}<0$. $R_{in}$ represents an upper
limit on CSL heating so that
\begin{equation}\label{eq:Rin}
R_{in}=\frac{C_{N}}{f_{T}}\left[\frac{f_{N}}{\tau_{N}}-\frac{f}{\tau_{f}}\right]-\frac{\mu_{N}}{\tau_{N}}+\frac{1}{\tau_{1}}\frac{U_{av}}{N}-\frac{1}{N}\left.\frac{dU}{dt}\right|_{cool}-\frac{1}{N}\left.\frac{dU}{dt}\right|_{TBR}
\end{equation}
and

\begin{equation}
\lambda<R_{in}\frac{4}{3A^{2}k_{B}T_{CSL}}.\label{Anal11}
\end{equation}

\section{Experimental Results}

Hanns-Christoph N\"agerl and Manfred Mark \cite{Mark} of the University
of Innsbruck have provided us with data for cesium condensates in
four different optical traps. These data were gathered from their ongoing
study of this system and the experiment was not designed with our
purposes in mind. Thus while the limit we get on $\lambda$ is rather
good, the result must be considered tentative,  serving as a
model for a more specific later experiment. Parameters of the traps are given in Table I:
\vspace{.2 in}

\begin{tabular}{|c|c|c|c|}
\hline 
Trap & freqs (Hz) & Depth (nK) & $a_{s}$ ($a_{0})$\tabularnewline
\hline 
\hline 
1 & 20.5, 22.0, 30.0 & 158 & 232\tabularnewline
\hline 
2 & 14.3, 15.5, 21.1 & 79 & 232\tabularnewline
\hline 
3 & 22, 23.5, 32 & 158 & 250\tabularnewline
\hline 
4 & 15.5, 16.4, 22.6 & 79 & 250\tabularnewline
\hline 
\end{tabular}

Table I.  Parameters for data for four cesium traps. The scattering length $a_{s}$ is given in units of the Bohr radius $a_{0}$.
\vspace{.2 in}

We shall illustrate the calculation of the right-hand side of (\ref{eq:Rin}) with data from Trap 1 and present the results
for the other three traps.

To begin, in order to calculate the contribution of $N^{-1}dU/dt$ (terms on the right-hand side of Eq.(\ref{eq:LHS})), we need 
$\tau_{N}$ and $\tau_{f}$.  We show below how we obtain these quantities, by fitting the $N(t)$ and $f(t)$ vs $t$ data with 
curves obtained by theoretical analysis of TBR and evaporative cooling. 

We also need  the values of $C$, $\mu_{N}$, $f_{T}$ and $f_{N}$.  These are obtained from
Hartree-Fock theory (as described by
Dalfovo et al \cite{Stringari-RMP}), and are given at the end of section IVD.  This approach has been found \cite{HKN}
to give results very close to Monte Carlo estimates. 

Following this, we shall  give a brief discussion of the estimates of the various energy
sources for cooling and heating.

\subsection{Evaluation of $\tau_{N}$ from TBR and evaporative cooling}

Three-body recombination in cesium has been extensively studied by
the Innsbruck group \cite{Weber,Kraemer,Haller}. The particle loss
rate due to TBR is \cite{Burt} 
\begin{equation}
\frac{dN(t)}{dt}=-L_{3}\int d\mathbf{r}n^{3}(\mathbf{r},t)\label{eq:Nrate}
\end{equation}
where $n(\mathbf{r},t)$ is the particle density and $L_{3}$ is the
rate of ejected particles; it is $3K_{3}$ where $K_{3}$ is the rate
at which triples form if we assume that the trap height is small enough
that all three particles are ejected. The parameter $K_{3}^{nc}$
for the non-condensate gas is 3! times larger than that for the condensate
$K_{3}^{c}$ because of exchange terms for differing states. 

When
there is both a condensate with density $n_{0}(\mathbf{r})$ and a
thermal cloud of density $n_{T}(\mathbf{r})$,  the result is 
\cite{KSS}
\begin{equation}\label{eq:TBR}
\frac{dN(t)}{dt}=-L_{3}^{c}\int d\mathbf{r}\left[n_{0}^{3}+9n_{0}^{2}n_{T}+18n_{0}n_{T}^{2}+6n_{T}^{3}\right]
\end{equation}
with $L_{3}^{c}=3K_{3}^{c}$. The values of $K_{3}^{c}$ for
cesium are given in Refs \cite{Kraemer,Haller}. Using the curves
in Ref. \cite{Haller} for pure cesium condensates we find, for $a_{s}=232a_{0}$,
the value $K_{3}\approx1.7\times10^{-40}$m$^{6}/$s. However for
this $a_{s}$ value, Ref. \cite{Kraemer}'s study of thermal cesium gases
finds a value about five times smaller. 

To compute the rate (\ref{eq:TBR}), we use the finite-temperature Thomas-Fermi approximation
for the condensate \cite{Stringari-RMP}: 
\begin{equation}
n_{0}(\mathbf{r})=\frac{1}{g}\left[\mu-V_{ext}({\bf r})\right]\theta(\mu-V_{ext}({\bf r}))\label{eq:CondDen}
\end{equation}
where $\theta(x)$ is the step function, $\mu$ is the chemical potential (\ref{HF1}), $V_{ext}$ is the harmonic oscillator potential and $g=4\pi\hbar^{2}a_{s}/m$. The  Hartree-Fock
approximation for the thermal gas is: 
\begin{equation}
n_{T}({\bf r})=\frac{1}{\lambda_{T}^{3}}g_{3/2}\left(\exp\left[-\beta\left(V_{ext}({\bf r})+2gn_{0}({\bf r})-\mu\right)\right]\right)\label{eq:ThermalDens}
\end{equation}
where the Bose integral is $g_{3/2}(z)=\sum_{l=1}^{\infty}z^{k}/k^{3/2}$, $\beta\equiv 1/k_{B}T$ and the thermal wavelength is $\lambda_{T} =\hbar \sqrt{2\pi/mk_{B}T}$.
We have neglected the interaction between condensate and thermal cloud
in $n_{0}$ and that between thermal atoms in $n_{T}$, so we do not
have to iterate the equations. 

\begin{figure}[h]
\centering \includegraphics[width=3in]{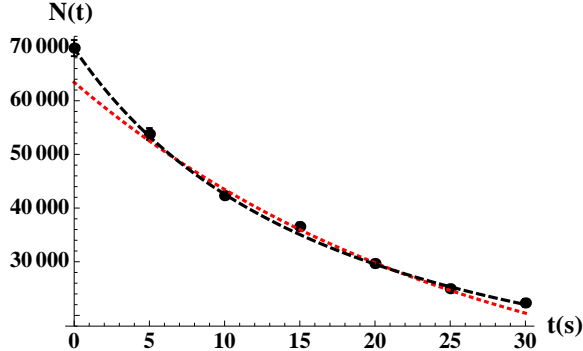}

\caption{Decay of particle number (black dots)  for a condensate plus
thermal gas  of Trap 1. The curve is fitted with a theory based on three-body recombination and evaporative cooling. We find parameters $K_{3}^{c}=3.6 \times10^{-41}$m$^{6}$/s
and $\tau_{cool}=62$ s. The fit also gives $N(0)=7.0 \times 10^{4}$ and $\tau_{N} =18\pm 0.5$ s.  A simple exponential fit is shown by the red dotted line.  }

\label{fig:NewTrap1Fit}
\end{figure}

It is to be expected  that Eq.(\ref{eq:TBR}) is not accurate at large times.  Then, 
 the density at the origin becomes small so TBR is diminished, and evaporative
cooling dominates. So we add a term $-N(t)/\tau_{cool}$ to the right hand side of (\ref{eq:TBR}) 
to account for that. Then, $N(t)$ is calculated and the best fit to the data is obtained by adjusting $K_{3}^{c}$, $N(0)$, and $\tau_{cool}$.   In fitting the data for $N(t)$ we set the temperature at a value determined from the initial condensate fraction $f(0)$ (see below) and make the approximation that it does not change during the decay. This process 
is shown in Fig. \ref{fig:NewTrap1Fit} for Trap 1 where we find $\tau_{N}=24$ s. See Table II for the parameters from all four traps.

The curve resulting from the TBR analysis plus the added exponential describing evaporative cooling appears to provide a considerably better fit to the data than the exponential
alone: with just the latter, the best fit yields $\tau_{N} =24$s.  Our value of
$K_{3}^{c}$ is smaller than those expected from Ref. \cite{Haller}
but is near that found in Ref. \cite{Kraemer}. 

We find similar results
for other traps provided by the Innsbruck group as shown in Table II.

\subsection{Evaluation of $\tau_{f}$}

\begin{figure}[h]
\centering \includegraphics[width=3in]{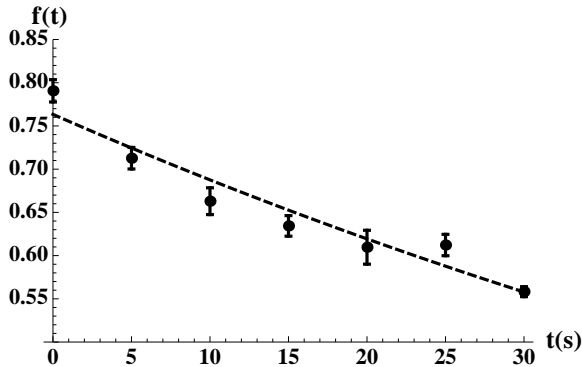}

\caption{Decay of condensate fraction $f(t)$ (black dots) for a condensate plus
thermal gas in Trap 1. An exponential best fit gives a decay constant $\tau_{f}=96$
s and $f(0)=0.76$. The latter parameter allows us to evaluate $T$. }

\label{fig:NewTrap2Cond}
\end{figure}

The other data we have to analyze is the condensate fraction. This
is fit with an exponential as shown in Fig. \ref{fig:NewTrap2Cond}. This is all that is needed. It is true that the condensate is the major contributor to the TBR losses because the TBR's dependence is on density to the third power and the condensate is much more dense than the cloud. However, as the TBR process takes place, evaporative cooling and other particle loss in the thermal cloud also takes place, and re-thermalization restores particles to the condensate, causing $\tau_{f}$ to be much larger than $\tau_{N}$. We fit these cumulative complex processes with an exponential. The exponential fit gives not only  $\tau_{f}$  but also $f(0)$ and from that $T$ via Eq. (\ref{HF3}). 

Treating each of the four traps data sets in this way we can get a
range of values for the parameters used in Eq.(\ref{eq:LHS}) to
evaluate the initial value of $(1/N)dU/dt$ and the heating and cooling
energy rates. We show the results in Table II.

\begin{tabular}{|c|c|c|c|c|c|c|c|c|}
\hline 
Trap  & $K_{3}$($10^{-41})$m$^{6}$/s & $\tau_{cool}$(s) & $\tau_{N}$  & $\tau_{f}$  & $f(0)$  & $T$  & $N(0)$(10$^{4}$)  & $T_{c}$(nK)\tabularnewline
\hline 
\hline 
1  & $3.6\pm0.4$ & $62\pm12$ & $17.5\pm0.5$  & $96\pm9$  & $0.76\pm0.01$  & $21\pm0.6$  & $7.0\pm0.2$  & $44$\tabularnewline
\hline 
2  & $2.1\pm0.6$ & $43\pm4$ & $28\pm1$  & $383\pm205$  & $0.75\pm0.02$  & $16\pm$0.5  & $8.3\pm0.1$  & $33$\tabularnewline
\hline 
3  & $3.2\pm0.1$ & $67\pm2$ & $17.2\pm0.1$  & $61\pm4$  & $0.71\pm0.02$  & $25\pm0.6$  & $8.2\pm0.1$  & $50$\tabularnewline
\hline 
4  & $3.5\pm0.4$ & $75\pm10$ & $24\pm1$  & $203\pm31$  & $0.82\pm0.01$  & $15\pm0.4$  & $9.0\pm0.3$  & $36$\tabularnewline
\hline 
\end{tabular}

$\mathbf{Caption}$: Table II. Ranges in fitted parameters to the
data in four traps of the Innsbruck group.

\subsection{ Energy sources}

Next we turn to the contributions  of the 
remaining three terms which make up $R_{in}$ in Eq.(\ref{eq:Rin}), evaporative cooling, TBR loss, and foreign atom collisions.  

\subsubsection{Evaporative cooling}

The LWR theory \cite{LRW} of evaporative cooling is consistent with
the more qualitative derivation of Pethick and Smith \cite{Pethick-book}.
While the expressions for $\tau_{cool}$ and $dU/dt|_{cool}$ given
by LWR are somewhat involved we find that  the
cooling power is accurately summarized by the simple formula
\begin{equation}
\frac{1}{N}\frac{dU}{dt}_{cool}=-\frac{\alpha E_{B}}{\tau_{cool}}\label{eq:dUdtcool}
\end{equation}
where $E_{B}$ is the energy to escape the trap (trap height energy  $\varepsilon_{w}$
minus zero-point energy) and $\alpha=1.12$. For Trap 1 the fitting
of the $N(t)$ vs $t$ data gives 
 $\tau_{cool}=35$s yielding the  rate $-2.8$ nK/s.
\subsubsection{TBR heating}

There are two references we know of that discuss the heating caused
by three-body recombination \cite{Weber,Werner}. These both apply
to thermal gases in the dilute (classical) limit where the kinetic
energy cancels out of the problem and makes them inappropriate for
our case of a mixed condensate and thermal gas. There is heating because
the density cubed factor favors recombination in the center of the
trap where the particles have lower energy. Thus, when these particles
are ejected, each has less energy than the average energy per particle
in the system. The excess energy left behind is shared among the remaining
particles during re-equilibrization and is a heating effect. The particles
near the center of the trap that are mostly involved in recombining
are the condensate particles. Even with the Thomas-Fermi condensate
wave function we have the condensate much nearer the center of the
trap than the thermal particles. Moreover, we are concerned with condensate
fractions of 0.70 to 0.80, which means the thermal density is small
anyway. Thus the majority of particles taking part in the three-body
recombination are condensate particles. However the thermal particles
have larger energies and contribute more to the average energy. Thus
we can estimate the energy lost per particle by TBR to be the energy
per particle at $T=0,$ $E_{cond}=U(T=0)/N=\eta k_{B}T_{c}/7$ using
Eq. (\ref{HF4}) for the interacting energies. This is smaller than the average
energy per particle $E_{av}=U(T)/N.$ So an estimate for the heating
rate is
\begin{equation}
R_{TBR}=(E_{av}-E_{cond})/\tau_{N}
\end{equation}
where, as before, $\tau_{N}$ is the initial particle number lifetime.
Calculations for Trap 1 give a value of 0.6 nK/s. 

\subsubsection{Foreign atom collisions and laser fluctuations}

Bali et al \cite{Bali} have estimated the collision rate between
various foreign and trapped alkali atoms for shallow traps as a function
of background gas pressure. The largest rate is due to Cs-Cs collisions.
We might assume that lost cesium atoms stay in an ``Oort cloud''
\cite{Cornell} and occasionally pass through the trapped gas. The
estimated background gas pressure in the experiments we are analyzing
is on the order of $2\times10^{-11}$mbar \cite{Mark} from which
we can get the density of the background gas at 300K. We find that
Cs-Cs collisions would occur at with a collision time of $\tau_{1}=320$
s. Collisions in which the trap atoms are not ejected from the trap
lead to heating, which from Ref. \cite{Bali} is on the order of $0.02$
nK/s, much too small to be relevant. If we assume all these collisions
cause atoms to be ejected, then the energy rate contribution can be
estimated as $-U_{av}/(N\tau_{1})$ where $U_{av}$ is the average
energy in the trapped gas. This then yields a rate on the order of
$-0.1$ nK/s, which is on the edge of being important. Of course if
the background gas pressure were, say, ten times larger, this
would be proportionately larger and be a major contributor. 

 Mark \cite{Mark} has done a study of the fluctuations in position
and intensity of the trapping lasers. Savard et al  \cite{Savard} shows that the laser positioning fluctuations give rise to a heating rate 
\begin{equation}
\frac{d\left\langle E\right\rangle }{dt}=\frac{\pi}{2}m\omega_{tr}^{4}S(\omega_{tr})
\end{equation}
where $\omega_{tr}$ is the trap frequency
and $S(\omega_{tr})$ is the power spectral density of the positioning fluctuations; these reach a maximum of about $\sqrt{S}=6 \times 10^{-3}$ $ \mu$m/$\sqrt{\mathrm{Hz}}$ corresponding to a negligible heating($<0.02$nK/s).

\subsection{Summary}

Table II shows the results from the four trap data sets. We use 
Eq.(\ref{eq:Rin}) to evaluate the unaccounted energy rate $R_{in}$,
our upper limit on CSL heating. In the table, $\frac{1}{N}\frac{dU}{dt}$ is given by  Eq.(\ref{eq:LHS})  and $W$,  the sum of all
the energy sources we have included in the computations, is the sum of the three columns before it. All energy
rates are per particle. 

\begin{tabular}{|c|c|c|c|c|c|c|}
\hline 
TRAP & $\frac{1}{N}\frac{dU}{dt}$(nK/s) & $\frac{1}{N}\frac{dU}{dt}_{cool}$ & $\frac{1}{N}\frac{dU}{dt}_{TBR}$ & $\frac{-U_{av}}{\tau_{1}N}$ & $W$ & $R_{in}=\frac{1}{N}\frac{dU}{dt}-W$\tabularnewline
\hline 
\hline 
1 & $-1.4\pm0.1$ & $-2.8\pm0.6$ & $0.57\pm0.06$ & $-0.08$ & $-2.3\pm0.7$ & $0.9\pm0.7$\tabularnewline
\hline 
2 & $-0.8\pm0.1$ & $-2.1\pm0.2$ & $0.28\pm0.02$ & $-0.06$ & $-1.9\pm0.2$ & $1.1\pm0.2$\tabularnewline
\hline 
3 & $-1.6\pm0.1$ & $-2.6\pm0.1$ & $0.84\pm0.03$ & $-0.10$ & $-1.9\pm0.1$ & $0.3\pm0.2$\tabularnewline
\hline 
4 & $-0.8\pm0.1$ & $-1.2\pm0.2$ & $0.23\pm0.02$ & $-0.05$ & $-1.0\pm$0.2 & $0.2\pm0.2$\tabularnewline
\hline 
\end{tabular}

$\mathbf{Caption}$: Table III. Data estimates of results needed to
evaluate $R_{in}$, the upper limit on the net energy input rate in
the four traps.

Thus we have an average of $R_{in}=0.6\pm0.5$ nK/s
per particle. Using Eq.(\ref{Anal11}) we get a limit on $\lambda$
of 
\begin{equation}
\lambda<1(\pm1)\times10^{-7}/\mathrm{s}
\end{equation}
which is bested only by Fu's limit $\lambda<1\times10^{-9}/\mathrm{s}$
at present\cite{Fu}. 

In conclusion, we have presented an analysis of the heating of a Bose Einstein condensate according to the CSL theory of dynamical collapse.  We have derived the relevant evolution of the density matrix, and  thereby obtained rate equations describing the evolution of the population of atoms occupying the various energy levels in a bound state. We then applied this to the specific problem of a BEC and its attendant thermal cloud. We considered the other processes which compete with CSL heating in altering state populations in a practical experiment. Using data on cesium BEC's kindly supplied by  Hanns-Christoph N\"agerl and Manfred Mark, we found an upper limit on the parameter $\lambda$ which governs the rate of collapse in the CSL theory.

Given the many uncertainties
in our calculations, our result should be regarded as provisional, to be improved by 
an experiment of this kind specifically tailored to obtain a more precise energy audit and so reduce $R_{in}$ and its uncertainty and improve the limit on $\lambda$. Features of such an explicitly designed experiment would include heavy atomic mass (like cesium), low background of foreign atoms, systematic measurement of three-body recombination or its elimination, and a sufficiently high barrier to eliminate evaporative cooling as much as possible. An attractive possibility is to use a box boundary  \cite{Hadz} , which can have the advantage of a uniformly low density, minimizing interactions and three-body recombination and lengthening the BEC lifetime.

\acknowledgments

We are extremely grateful to Manfred Mark and Hanns-Christoph N\"agerl for their permission to use unpublished data for our analysis.  We would also like to thank David Hall and Fabrice Gerbier for many interesting and useful discussions.

\appendix

\section{Energy Increase}

  If the energy of the $i$th state is 
$\epsilon_{i}\equiv \langle \varphi_{i}|H|\varphi_{i}\rangle$, so $H={\bf P}^{2}/2m +V({\bf Y})=\sum_{i}\epsilon_{i}|\varphi_{i}\rangle\langle\varphi_{i}|$, the expression for  the rate of increase of the ensemble-averaged energy, $\overline{E}\equiv\sum_{i}\epsilon_{i}\langle N_{i}\rangle$ follows from the rate equations (\ref{19-FL}), 
(\ref{20-FL}):
\begin{eqnarray}\label{Rate'2}
\frac{d}{dt}\overline{E}&=&-\lambda A^{2}\overline{E}+\lambda A^{2}\int d\mathbf{y}\int d\mathbf{y}'
e^{-(\mathbf{y}-\mathbf{y}')^{2}/4a^{2}} 
\sum_{i}\epsilon_{i}\varphi_{i}(\mathbf{y})\varphi_{i}^{*}(\mathbf{y}')
\sum_{k}\varphi_{k}^{*}(\mathbf{y})\varphi_{k}(\mathbf{y}')\langle N_{k}\rangle.\nonumber\\
\end{eqnarray}
\noindent The sum over $i$ may be expressed as a matrix element of $H$, and then the exponential also can be expressed in terms of position operators: 
\begin{eqnarray}\label{Rate'3}
\frac{d}{dt}\overline{E}&=&-\lambda A^{2}\overline{E}+\lambda A^{2}\int d\mathbf{y}\int d\mathbf{y}'
e^{-(\mathbf{y}-\mathbf{y}')^{2}/4a^{2}}\langle {\bf y}|H| {\bf y}'\rangle
\sum_{k}\varphi_{k}^{*}(\mathbf{y})\varphi_{k}(\mathbf{y}')\langle N_{k}\rangle\nonumber\\
&=&-\lambda A^{2}\overline{E}+\lambda A^{2}\int d\mathbf{y}\int d\mathbf{y}'
\langle {\bf y}|e^{-(\mathbf{Y}_{L}-\mathbf{Y}_{R})^{2}/4a^{2}}H| {\bf y}'\rangle
\sum_{k}\varphi_{k}^{*}(\mathbf{y})\varphi_{k}(\mathbf{y}')\langle N_{k}\rangle.
\end{eqnarray}
\noindent Expanding the exponential, the commutator with $H$ has only the non-vanishing part
\begin{eqnarray}\label{Rate'4}
\frac{d}{dt}\overline{E}&=&-\lambda A^{2}\overline{E}+\lambda A^{2}\int d\mathbf{y}\int d\mathbf{y}'
\langle {\bf y}|\{H-\frac{1}{4a^{2}}[{\bf Y},[{\bf Y},H]]\}| {\bf y}'\rangle
\sum_{k}\varphi_{k}^{*}(\mathbf{y})\varphi_{k}(\mathbf{y}')\langle N_{k}\rangle.\nonumber\\
\end{eqnarray}
\noindent The first term in the curly bracket cancels $-\lambda A^{2}\overline{E}$, since
\[
\int d\mathbf{y}\int d\mathbf{y}'
\langle {\bf y}|H| {\bf y}'\rangle
\sum_{k}\varphi_{k}^{*}(\mathbf{y})\varphi_{k}(\mathbf{y}')\langle N_{k}\rangle=\sum_{k}\langle\varphi_{k}|H|\varphi_{k}\rangle\langle N_{k}\rangle
=\sum_{k}\epsilon_{k}\langle N_{k}\rangle=\overline{E}.
\]
\noindent The commutator can readily be evaluated, with the result
\begin{eqnarray}\label{Rate'5}
\frac{d}{dt}\overline{E}&=&-\lambda A^{2}\frac{1}{4a^{2}}\int d\mathbf{y}\int d\mathbf{y}'
\langle {\bf y}|[{\bf Y},[{\bf Y},\frac{{\bf P}^{2}}{2m}]]| {\bf y}'\rangle
\sum_{k}\varphi_{k}^{*}(\mathbf{y})\varphi_{k}(\mathbf{y}')\langle N_{k}\rangle\nonumber\\
&=&\lambda A^{2}\frac{1}{4a^{2}}\frac{3}{2m}\int d\mathbf{y}\int d\mathbf{y}'
\delta({\bf y}-{\bf y}')\sum_{k}\varphi_{k}^{*}(\mathbf{y})\varphi_{k}(\mathbf{y}')\langle N_{k}\rangle=\lambda A^{2}N\frac{3}{4ma^{2}}.  
\end{eqnarray}
\noindent using  $ \int d\mathbf{y}| \varphi_{k}(\mathbf{y})|^{2}=1$ and $\sum_{k}\langle N_{k}\rangle=N$.  
This linear increase of  $\overline{E}$ with $t$ 
 is a well-known\cite{Reviews} consequence of CSL. It is often expressed as 
 \begin{equation}\label{Rate'6}
\frac{d}{dt}\overline{E}=\lambda \frac{3}{4Ma^{2}}\frac{\cal M}{M}
\end{equation}
\noindent where ${\cal M}$ is the total mass of all the atoms, and $M$ is the mass of a nucleon, which follows from  Eq.(\ref{Rate'5}) with use of $m=AM$ and 
${\cal M}=MAN$.

\section{Rate Equations for Box and Harmonic Oscillator}\label{AppA}

For both a box potential with side length $\sigma$ and the harmonic oscillator with $m\omega=1/\sigma^{2}$, starting from the rate equations Eq.(\ref{19-FL}), (\ref{20-FL}):

\begin{equation}\label{AppA1}
\frac{d}{dt}\langle N_{i}\rangle=-\lambda A^{2}\langle N_{i}\rangle+\lambda
A^{2}\sum_{k}  \int d\mathbf{y}\int d\mathbf{y}'e^{-\frac{1}{4a^{2}}
(\mathbf{y}-\mathbf{y}^{\prime}) ^{2}}  
\varphi_{i}(\mathbf{y})\varphi_{i}^{*}(\mathbf{y}')
\varphi_{k}^{*}(\mathbf{y})\varphi_{k}(\mathbf{y}')       \langle N_{k}\rangle,
\end{equation}
replace the state label $i$ by the indices $n_{1}n_{2}n_{3}$ to characterize the states $\phi_{n_{1}}(x)\phi_{n_{2}}(y)\phi_{n_{3}}(z)$, obtaining:
\begin{equation}\label{AppA2}
\frac{d}{dt}\langle N_{n_{1}n_{2}n_{3}}\rangle=-\lambda A^{2}\langle N_{n_{1}n_{2}n_{3}}\rangle+
\lambda A^{2}\sum_{m_{1}m_{2}m_{3}}I_{n_{1}m_{1}}I_{n_{2}m_{2}}I_{n_{3}m_{3}}\langle N_{m_{1}m_{2}m_{3}}\rangle
\end{equation}
\noindent where
\begin{equation}\label{AppA3}
I_{nm}\equiv \int dy\int dy'\phi_{n}(y)\phi_{n}( y')\phi_{m}(y)\phi_{m}( y')e^{-\frac{1}{4a^{2}}[y-y']^{2}}.
\end{equation}
\subsection {Box}\label{A}
For the box,  Eq.(\ref{AppA3}) becomes:
\begin{eqnarray}\label{A1}
I_{nm}
&=&\Big(\frac{2}{\sigma}\Big)^{2}\int_{0}^{\sigma} dy\int _{0}^{\sigma}dy' \sin\frac{n\pi y}{\sigma}\sin\frac{n\pi y'}{\sigma}\sin\frac{m\pi y}{\sigma}\sin\frac{m\pi y'}{\sigma}e^{-\frac{1}{4a^{2}}[y-y']^{2}}\nonumber\\
&=&\Big(\frac{2}{\pi}\Big)^{2}\int dx\int dx' \sin nx\sin nx'\sin mx\sin mx'e^{-\frac{1}{4\alpha'^{2}}[x-x']^{2}}.
\end{eqnarray}
\noindent where $\alpha'\equiv\pi(a/\sigma)$.
\noindent  Because the exponential is large only for small $|x-x'|\lesssim 2\alpha'$, we make the approximation $\sin nx\sin nx'=(1/2)[\cos n(x+x')+\cos n(x-x')]\approx(1/2)\cos n(x-x')$,  obtaining
\begin{eqnarray}\label{A2}
I_{nm}
&\approx&\Big(\frac{1}{\pi}\Big)^{2}\int_{0}^{\pi} dx\int_{0}^{\pi} dx' \cos n(x-x')\cos m(x-x')e^{-\frac{1}{4\alpha'^{2}}[x-x']^{2}}\nonumber\\
&=&\frac{1}{2\pi^{2}}\int_{-\pi}^{\pi} dv\cos nv\cos mv e^{-\frac{1}{4\alpha'^{2}}v^{2}}\int_{v}^{2\pi-v}du\nonumber\\
&\approx&\frac{1}{\pi}\int_{-\pi}^{\pi} dv\cos nv\cos mv e^{-\frac{1}{4\alpha'^{2}}v^{2}}\nonumber\\
&=&\frac{1}{2\pi}\int_{-\infty}^{\infty} dv[\cos (n-m)v+\cos (n+m)v] e^{-\frac{1}{4\alpha'^{2}}v^{2}}\nonumber\\
&=&\frac{\alpha'}{\sqrt{\pi}}\Big[e^{-(n-m)^{2}\alpha'^{2}}+e^{-(n+m)^{2}\alpha'^{2}}\Big].
\end{eqnarray}
\noindent where in the second line we have changed variables to $v\equiv x-x' $, $u\equiv x+x' $, and in the third line we have used  $\alpha'<<\pi$.   

Putting Eq.(\ref{A2}) into Eq.(\ref{AppA2}), we arrive at:
\begin{eqnarray}\label{A2.5}
\frac{d}{dt}\langle N_{n_{1}n_{2}n_{3}}\rangle=-\lambda A^{2}\langle N_{n_{1}n_{2}n_{3}}\rangle+\lambda A^{2}\frac{\alpha'^{3}}{(\pi)^{3/2} }\sum_{m_{1}m_{2}m_{3}}
\prod_{k=1}^{3}\Big[e^{-\alpha'^{2}(n_{k}-m_{k})^{2}}+e^{-\alpha'^{2}(n_{k}+m_{k})^{2}}\Big]\langle N_{m_{1}m_{2}m_{3}}\rangle.\nonumber\\
\end{eqnarray} 

What we would like, however, are rate equations for the particle number in a state of given energy, i.e., for $\langle N_{n}\rangle\equiv\sum_{n_{1}n_{2}n_{3}}\langle N_{n_{1}n_{2}n_{3}}\rangle$ where the sum is over all $n_{i}$ such that $n=\sqrt{n_{1}^{2}+n_{2}^{2}+n_{3}^{2}}$ for the box (and 
$n=n_{1}+n_{2}+n_{3}$ for the harmonic oscillator).  So, we evaluate Eq.(\ref{A2.5}) summed over all $n_{i}$ subject to the condition  $n=[n_{1}^{2}+n_{2}^{2}+n_{3}^{2}]^{1/2}$:
\begin{eqnarray}\label{A3}
S_{n}&\equiv&\frac{\alpha'^{3}}{(\pi)^{3/2} }\sum_{n_{1}n_{2}n_{3}}\sum_{m_{1}m_{2}m_{3}}
\prod_{k=1}^{3}\Big[e^{-\alpha'^{2}(n_{k}-m_{k})^{2}}+e^{-\alpha'^{2}(n_{k}+m_{k})^{2}}\Big]\langle N_{m_{1}m_{2}m_{3}}\rangle.
\end{eqnarray}

Setting  $m\equiv [m_{1}^{2}+m_{2}^{2}+m_{3}^{2}]^{1/2}$, we approximate the sum over $n_{i}$ by an integral, obtaining: 
\begin{subequations}
\begin{eqnarray}\label{A4}
S_{n}&\equiv&\frac{\alpha'^{3}}{(\pi)^{3/2}}\sum_{m_{1}m_{2}m_{3}}e^{-\alpha'^{2}[n^{2}+m^{2}]}
J_{m_{1}m_{2}m_{3}}\langle N_{m_{1}m_{2}m_{3}}\rangle \label{A4a}\\
 J_{m_{1}m_{2}m_{3}}&\equiv &\int_{0}^{n} dn_{1}dn_{2}dn_{3}\delta\Big[n-(n_{1}^{2}+n_{2}^{2}+n_{3}^{2})^{1/2}\Big]
 \prod_{k=1}^{3}\Big[e^{2\alpha'^{2}n_{k}m_{k}}+e^{-2\alpha'^{2}n_{k}m_{k}}\Big]\label{A4b}
 \end{eqnarray}
\end{subequations}

 To evaluate $ J_{m_{1}m_{2}m_{3}}$, we switch to polar coordinates:
 \begin{subequations}
\begin{eqnarray}\label{A5}
 J_{m_{1}m_{2}m_{3}}&= &\int_{0}^{n}r^{2} dr\delta(n-r)\int_{0}^{\pi/2}d\theta \sin\theta\int_{0}^{\pi/2}d\phi\nonumber\\
 &&\negthinspace\negthinspace\negthinspace\negthinspace\negthinspace\negthinspace\negthinspace\negthinspace
 \cdot8\cosh(2\alpha'^{2}nm_{3}\cos\theta)\cosh(2\alpha'^{2}nm_{1}\sin\theta\cos\phi)\cosh(2\alpha'^{2}nm_{2}\sin\theta\sin\phi)\label{A5a}\\
 &=& n^{2} \int_{0}^{\pi}d\theta \sin\theta\int_{0}^{2\pi}d\phi\nonumber\\
 &&\negthinspace\negthinspace\negthinspace\negthinspace\negthinspace\negthinspace\cdot\cosh(2\alpha'^{2}nm_{3}\cos\theta)\cosh(2\alpha'^{2}nm_{1}\sin\theta\cos\phi)\cosh(2\alpha'^{2}nm_{2}\sin\theta\sin\phi)\label{A5b}
  \end{eqnarray}
\end{subequations}
 \noindent The integral in Eq.(\ref{A5a}) is over the first quadrant, but  the integral in Eq.(\ref{A5b}) is over all eight quadrants since, for the integral in any other quadrant,  the $\sin$'s and $\cos$'s change sign, but the  $\cosh$'s do not change.  Upon writing $m_{i}=sv_{i}$, where ${\bf v}\cdot{\bf v}=1$, we see that
the product of the $\cosh$'s is the sum of 8 terms, each of the form $\exp 2\alpha'^{2}nm{\bf v}\cdot{\bf i}$, where ${\bf i}$ is a unit vector with (depending upon the term) components $\pm\sin\theta\cos\phi, \pm\sin\theta\sin\phi,\pm\cos\theta$.  Since the integral is over the whole solid angle, we may in each case rotate
the coordinate system, obtaining identical integrals for each, and thus
\begin{equation}\label{A6}
 J_{m_{1}m_{2}m_{3}}= 2\pi n^{2} \int_{-1}^{1}d\cos\theta e^{2\alpha'^{2}nm_{1}\cos\theta}= \frac{n\pi}{m\alpha'^{2}}\big[ e^{2\alpha'^{2}nm} - e^{-2\alpha'^{2}nm}  \big]
   \end{equation}
Putting Eq.(\ref{A6}) into Eq.(\ref{A4a}), and replacing the sum over $m$ by an integral, gives the result:
\begin{equation}\label{A7}
S_{n}= \frac{\alpha' n}{\sqrt{\pi}}\int_{0}^{\infty}\frac{dm}{m}\Big[e^{-\alpha'^{2}(n-m)^{2}}-e^{-\alpha'^{2}(n+m)^{2}}\Big]\langle N_{m}\rangle.
\end{equation}
Putting this into Eq.(\ref{A2.5}), we get the rate equations 
\begin{eqnarray}\label{A8}
\frac{d}{dt}\langle N_{n}\rangle=-\lambda A^{2}\langle N_{n}\rangle+\lambda A^{2}\frac{\alpha' n}{\sqrt{\pi}}\int_{0}^{\infty}\frac{dm}{m}\Big[e^{-\alpha'^{2}(n-m)^{2}}-e^{-\alpha'^{2}(n+m)^{2}}\Big]\langle N_{m}\rangle.
\end{eqnarray} 

Finally, we wish to express this as  rate equations for the energy density of states $\langle N_{\epsilon}\rangle=\langle N_{n}\rangle dn/d\epsilon$.  Using 
$n=\pi^{-1}\sigma \sqrt{2m\epsilon}$, we obtain
\begin{eqnarray}\label{A9}
\frac{d}{dt}\langle N_{\epsilon}\rangle&=&-\lambda A^{2}\langle N_{\epsilon}\rangle+\frac{\lambda A^{2}}{\sqrt{\pi k_{B}T_{CSL}/2}}\int_{0}^{\infty}d\sqrt{\epsilon'}\Big[e^{-\frac{2}{k_{B}T_{CSL}}(\sqrt{\epsilon}-\sqrt{\epsilon'})^{2}}-e^{-\frac{2}{k_{B}T_{CSL}}(\sqrt{\epsilon}+\sqrt{\epsilon'})^{2}}\Big]\langle N_{\epsilon'}\rangle.\nonumber\\
\end{eqnarray}

\subsection{Harmonic Oscillator}\label{B}
For the harmonic oscillator,  Eq.(\ref{AppA3}) becomes: 
\begin{eqnarray}\label{B1}
I_{nm}&\equiv& \int dy\int dy'\phi_{n}(y)\phi_{n}( y')\phi_{m}(y)\phi_{m}( y')e^{-\frac{1}{4a^{2}}[y-y']^{2}}\nonumber\\
&=&\frac{1}{\pi\sigma^{2}2^{n}2^{m}n!m!}\int dy\int dy' e^{-\frac{y^{2}}{\sigma^{2}}}e^{-\frac{y'^{2}}{\sigma^{2}}} H_{n}(y/\sigma)H_{n}(y'/\sigma)H_{m}(y/\sigma)H_{m}(y'/\sigma)e^{-\frac{1}{4a^{2}}[y-y']^{2}}\nonumber\\
&=&\frac{1}{\pi2^{n}2^{m}n!m!}\int dx\int dx' e^{-x^{2}}e^{-x'^{2}}H_{n}(x)H_{n}(x')H_{m}(x)H_{m}(x')e^{-\frac{1}{4\alpha^{2}}[x-x']^{2}},
\end{eqnarray}
\noindent where $\alpha\equiv a/\sigma<<1$. We shall use the asymptotic expression
\begin{equation}\label{B2}
H_{n}(x)\rightarrow e^{\frac{x^{2}}{2}}\sqrt{2^{n}n!}\Big(\frac{2}{\pi n}\Big)^{1/4}\cos\Big(x\sqrt{2n}-n\frac{\pi}{2}\Big)
\end{equation}
(good to near the turning points at $x\approx\pm\sqrt{2n}$, and a fairly good approximation even for relatively small values of $n$) in Eq.(\ref{B1}):
\begin{subequations}
\begin{eqnarray}\label{B3}
I_{nm}
&\approx &\frac{2}{\pi^{2}\sqrt{nm}}\int dx\int dx' \cos\Big(x\sqrt{2n}-n\frac{\pi}{2}\Big)\cos\Big(x'\sqrt{2n}-n\frac{\pi}{2}\Big)
\nonumber\\
&&\qquad\qquad\qquad\qquad\cdot\cos\Big(x\sqrt{2m}-m\frac{\pi}{2}\Big)\cos\Big(x'\sqrt{2m}-m\frac{\pi}{2}\Big)e^{-\frac{1}{4\alpha^{2}}[x-x']^{2}}
\label{B3a}\\
&\approx &\frac{1}{4\pi^{2}\sqrt{nm}}\int _{-c2\sqrt{2m}}^{c2\sqrt{2m}}du\int dv\cos\Big(v\sqrt{2n}\Big)\cos\Big(v\sqrt{2m}\Big)e^{-\frac{1}{4\alpha^{2}}v^{2}}\label{B3b}\\
&= &\alpha\frac{1}{\sqrt{2\pi n}}\Big[e^{-2\alpha^{2}(\sqrt{n}-\sqrt{m})^{2}} + e^{-2\alpha^{2}(\sqrt{n}+\sqrt{m})^{2}}  \Big]\label{B3c}
\end{eqnarray}
\end{subequations}
\noindent In Eq.(\ref{B3b}), beside changing variables to $u\equiv x+x'$, $u\equiv x-x'$  and discarding the negligible oscillating cos terms which depend upon $u$ as in Appendix A, we have put in limits on the
$u=x+x'$ variable.  This is because the approximation (\ref{B2}) is good only out to near the turning points $\pm\sqrt{2m}$, beyond which the $H_{m}$'s decay exponentially and give a negligible contribution.  The constant $c$ is determined by the requirement that $\int_{0}^{\infty} dn I_{nm}=1$ (so particle number is constant), and is found  to be $c=\pi/2$.  

Putting Eq.(\ref{B3c}) into Eq.(\ref{AppA2}), we arrive at:
\begin{eqnarray}\label{B2.5}
\frac{d}{dt}\langle N_{n_{1}n_{2}n_{3}}\rangle=-\lambda A^{2}\langle N_{n_{1}n_{2}n_{3}}\rangle+\lambda A^{2}\frac{\alpha^{3}}{(\pi)^{3/2} }\sum_{m_{1}m_{2}m_{3}}
\prod_{k=1}^{3}\Big[e^{-2\alpha^{2}(n_{k}-m_{k})^{2}}+e^{-2\alpha^{2}(n_{k}+m_{k})^{2}}\Big]\langle N_{m_{1}m_{2}m_{3}}\rangle.\nonumber\\
\end{eqnarray} 

Next required is that we evaluate Eq.(\ref{B2.5}) summed over all $n_{i}$ subject to the condition  $n=n_{1}+n_{2}+n_{3}$:
\begin{eqnarray}\label{B4}
S_{n}&\equiv&\frac{\alpha^{3}}{(2\pi)^{3/2} }\sum_{n_{1}n_{2}n_{3}}\sum_{m_{1}m_{2}m_{3}}\frac{1}{\sqrt{n_{1}n_{2}n_{3}}}
\prod_{k=1}^{3}\Big[e^{-2\alpha^{2}(\sqrt{n_{k}}-\sqrt{m_{k}})^{2}} + e^{-2\alpha^{2}(\sqrt{n_{k}}+\sqrt{m_{k}})^{2}}  \Big]\langle N_{m_{1}m_{2}m_{3}}\rangle.\nonumber\\
\end{eqnarray}
\noindent  Setting $m\equiv m_{1}+m_{2}+m_{3}$, and summing over all $n_{i}$ corresponding to $n$, with the sum approximated by an integral, we have
\begin{subequations}
\begin{eqnarray}\label{B5}
S_{n}&\equiv& \frac{\alpha^{3}}{(2\pi)^{3/2}}\sum_{m_{1}m_{2}m_{3}}e^{-2\alpha^{2}[n+m]}
J_{m_{1}m_{2}m_{3}}\langle N_{m_{1}m_{2}m_{3}}\rangle \label{B5a}\\
 J_{m_{1}m_{2}m_{3}}&\equiv &\int_{0}^{p}\frac{ dn_{1}dn_{2}dn_{3}}{\sqrt{n_{1}n_{2}n_{3}}}\delta(n-n_{1}-n_{2}-n_{3})
 \prod_{k=1}^{3}\Big[e^{4\alpha^{2}\sqrt{n_{k}m_{k}}}+e^{-4\alpha^{2}\sqrt{n_{k}m_{k}}}\Big] .                   \label{B5b}
 \end{eqnarray}
\end{subequations}

 To evaluate $ J_{m_{1}m_{2}m_{3}}$, we first set $n_{k}=x_{k}^{2}$ and then switch to polar coordinates:
\begin{subequations}
\begin{eqnarray}\label{B6}
 J_{m_{1}m_{2}m_{3}}&= &8\int_{0}^{\sqrt{n}} dx_{1}dx_{2}dx_{3}
 \delta(n-x_{1}^{2}-x_{2}^{2}-x_{3}^{2})
  \prod_{k=1}^{3}\Big[e^{4\alpha^{2}x_{k}\sqrt{m_{k}}}+e^{-4\alpha^{2}x_{k}\sqrt {m_{k}}}\Big]\label{B6a}\\
  &= &8\int_{0}^{\infty}r^{2}dr \delta(n-r^{2})\int_{0}^{\pi/2}d\theta \sin\theta  \int_{0}^{\pi/2}d\phi 8\cosh[4\alpha^{2}\sqrt{nm_{3}}\cos\theta]\nonumber\\
  &&\qquad\qquad\qquad\cdot\cosh[4\alpha^{2}\sqrt{nm_{1}}\sin\theta\cos\phi]\cosh[4\alpha^{2}\sqrt{nm_{2}}\sin\theta\sin\phi]\label{B6b}\\
   &= &4\sqrt{n}\int_{0}^{\pi}d\theta \sin\theta  \int_{0}^{2\pi}d\phi \cosh[4\alpha^{2}\sqrt{nm_{3}}\cos\theta]\nonumber\\
  &&\qquad\qquad\qquad\cdot\cosh[4\alpha^{2}\sqrt{nm_{1}}\sin\theta\cos\phi]\cosh[4\alpha^{2}\sqrt{nm_{2}}\sin\theta\sin\phi]\label{B6c}
 \end{eqnarray}
\end{subequations}
\noindent The integral in Eq.(\ref{B6b}) is over the first quadrant, but  the integral in Eq.(\ref{B6c}) is over all quadrants since, for the integral in any other quadrant,  the $\sin$'s and $\cos$'s change sign, but the  $\cosh$'s do not change.  Upon writing $m_{i}=mv_{i}^{2}$, where ${\bf v}\cdot{\bf v}=1$, we see that
the product of the $\cosh$'s is the sum of 8 terms, each of the form $\exp 4\alpha^{2}\sqrt{nm}{\bf v}\cdot{\bf i}$, where ${\bf i}$ is a unit vector with (depending upon the term) components $nm\sin\theta\cos\phi, nm\sin\theta\sin\phi,nm\cos\theta$.  Since the integral is over the whole solid angle, we may in each case rotate
the coordinate system, obtaining
\begin{eqnarray}\label{B7}
 J_{m_{1}m_{2}m_{3}}&= & 8\pi\sqrt{n}\int_{-1}^{1}d\cos\theta e^{4\alpha^{2}\sqrt{nm}\cos\theta }=\frac{2\pi}{\alpha^{2}\sqrt{m}}\Big [e^{4\alpha^{2}\sqrt{nm} }-
  e^{-4\alpha^{2}\sqrt{nm} }\Big].
 \end{eqnarray}

Putting Eq.(\ref{B7}) into Eq.(\ref{B5a}) gives the result:
\begin{equation}\label{B8}
S_{n}=\frac{\alpha}{\sqrt{2\pi}}\int_{0}^{\infty}\frac{dm}{\sqrt{m}}\Big[e^{-2\alpha^{2}(\sqrt{n}-\sqrt{m})^{2}}-e^{-2\alpha^{2}(\sqrt{n}+\sqrt{m})^{2}}\Big]\langle N_{m}\rangle. 
\end{equation}

Putting this into Eq.(\ref{B2.5}), we get the rate equations 
\begin{eqnarray}\label{B9}
\frac{d}{dt}\langle N_{n}\rangle=-\lambda A^{2}\langle N_{n}\rangle+\lambda A^{2}\frac{\alpha}{\sqrt{2\pi}}\int_{0}^{\infty}\frac{dm}{\sqrt{m}}\Big[e^{-2\alpha^{2}(\sqrt{n}-\sqrt{m})^{2}}-e^{-2\alpha^{2}(\sqrt{n}+\sqrt{m})^{2}}\Big]\langle N_{m}\rangle\end{eqnarray} 

Finally, we wish to express this as  rate equations for the energy density of states $\langle N_{\epsilon}\rangle=\langle N_{n}\rangle dn/d\epsilon$.  Using 
$n=\epsilon/\omega=\epsilon m\sigma^{2}$, we obtain
\begin{eqnarray}\label{B10}
\frac{d}{dt}\langle N_{\epsilon}\rangle&=&-\lambda A^{2}\langle N_{\epsilon}\rangle+\frac{\lambda A^{2}}{\sqrt{\pi k_{B}T_{CSL}/2}}\int_{0}^{\infty}d\sqrt{\epsilon'}\Big[e^{-\frac{2}{k_{B}T_{CSL}}(\sqrt{\epsilon}-\sqrt{\epsilon'})^{2}}-e^{-\frac{2}{k_{B}T_{CSL}}(\sqrt{\epsilon}+\sqrt{\epsilon'})^{2}}\Big]\langle N_{\epsilon'}\rangle.\nonumber\\
\end{eqnarray}

\section{Solution of Rate Equations }

The rate equations (\ref{B10}) and (\ref{A9}) are identical, despite their quite different potentials.  This is because, in both cases, the bound state wave functions are sinusoids of fixed wave-number, or well approximated by sinusoids.  That is a good approximation if  the potential energy is negligible compared to the kinetic energy for most of the range of $x$ between the classical turning points.  Since the natural energy range for CSL excitation is $k_{B}T_{CSL}$, which corresponds to sinusoid wavelengths of order $a$,  these rate equations and the solution below hold well for a wide range of momenta around $k=2\pi/a$ and larger, for any trap where the potential energy is 
$<<k_{B}T_{CSL}$ for most of the range of $x$.  This is true for the traps used in BEC experiments.

The solution of this rate equation can be found as follows. Set $\sqrt{\epsilon}\equiv z$,  $\sqrt{\epsilon'}\equiv z'$ in (\ref{B10}) or (\ref{A9}), and assume that 
$\langle N_{\epsilon'}\rangle$ is an antisymmetric function of $z'$ (where it has not been previously defined) so that  (\ref{B10}) or (\ref{A9}) may be written as
\begin{eqnarray}\label{C0}
\frac{d}{dt}\langle N_{z}\rangle&=&-\lambda A^{2}\langle N_{z}\rangle+\frac{\lambda A^{2}}{\sqrt{\pi k_{B}T_{CSL}/2}}\int_{-\infty}^{\infty}dz'e^{-\frac{2}{k_{B}T_{CSL}}(z-z')^{2}}\langle N_{z'}\rangle.
\end{eqnarray}

Apply the Fourier transform $g(k)\equiv\int_{-\infty}^{\infty}dze^{-ikz}\langle N_{z}\rangle$ to Eq.(\ref{C0}), with the result

\begin{equation}\label{C2}
\frac{d}{dt}g(k)=-\lambda A^{2}g(k)\Big[1-e^{-\frac{k^{2}k_{B}T_{CSL}}{8}}\Big]. 
\end{equation}
Solving for $g(k)$, and inverting the fourier transform, we obtain the solution to the rate equation, which can be written various ways:
\begin{subequations}
\begin{eqnarray}\label{C3}
 \langle N_{\epsilon}\rangle(t)&=& \frac{1}{2\pi}e^{-\lambda A^{2}t}\int_{0}^{\infty}d\sqrt{\epsilon'}\langle N_{\epsilon'}\rangle(0)\int_{-\infty}^{\infty} dk
 \Big[e^{ik(\sqrt{\epsilon}-\sqrt{\epsilon'})}-e^{ik(\sqrt{\epsilon}+\sqrt{\epsilon'})}\Big]
e^{\lambda A^{2}te^{-\frac{k^{2}k_{B}T_{CSL}}{8}}}\label{C3a}\\
&=& \langle N_{\epsilon}\rangle(0)e^{-\lambda A^{2}t}\nonumber\\
&&\negmedspace\negmedspace\negmedspace\negmedspace\negmedspace\negmedspace\negmedspace\negmedspace
\negmedspace\negmedspace\negmedspace\negmedspace\negmedspace\negmedspace\negmedspace\negmedspace
\negmedspace\negmedspace\negmedspace\negmedspace\negmedspace\negmedspace\negmedspace\negmedspace
+ \frac{1}{2\pi}e^{-\lambda A^{2}t}\int_{0}^{\infty}d\sqrt{\epsilon'}\langle N_{\epsilon'}\rangle(0)\int_{-\infty}^{\infty} dk
 \Big[e^{ik(\sqrt{\epsilon}-\sqrt{\epsilon'})}-e^{ik(\sqrt{\epsilon}+\sqrt{\epsilon'})}\Big]
\Big[e^{\lambda A^{2}te^{-\frac{k^{2}k_{B}T_{CSL}}{8}}}-1\Big]\label{C3b}\\
&=& \langle N_{\epsilon}\rangle(0)e^{-\lambda A^{2}t}\nonumber\\
&+&\frac{1}{\sqrt{\pi k_{B}T_{CSL}/2}}\int_{0}^{\infty}d\sqrt{\epsilon'}\langle N_{\epsilon'}\rangle(0)\sum_{s=1}^{\infty}e^{-\lambda A^{2}t}\frac{(\lambda A^{2}t)^{s}}{s!\sqrt{s}}
\Big[e^{-\frac{2}{sk_{B}T_{CSL}}(\sqrt{\epsilon}-\sqrt{\epsilon'})^{2}}-e^{-\frac{2}{sk_{B}T_{CSL}}(\sqrt{\epsilon}+\sqrt{\epsilon'})^{2}}\Big].\nonumber\\\label{C3c}
\end{eqnarray}
\end{subequations}

\subsection{Conservation Laws}
We now show that  constant particle number  and the proper linear energy increase are consequences of Eq.(\ref{C3c}): this supports the 
validity of the approximations made in obtaining   Eq.(\ref{A9}) or  Eq.(\ref{B10}).

Multiply Eq.(\ref{C3c})  by an arbitrary function $f(\epsilon)$, and integrate over all $\epsilon$ from 0 to $\infty$.  
Define $z=\sqrt{\epsilon}$  and, using $d\epsilon=2zdz$, one gets 
\begin{eqnarray}\label{CL0}
\overline f(t)&=& \overline f(0)e^{-\lambda A^{2}t}\nonumber\\
&&\negmedspace\negmedspace\negmedspace\negmedspace\negmedspace\negmedspace\negmedspace\negmedspace\negmedspace\negmedspace
\negmedspace\negmedspace\negmedspace\negmedspace\negmedspace\negmedspace
+\frac{1}{\sqrt{\pi k_{B}T_{CSL}/2}}\int_{0}^{\infty}d\sqrt{\epsilon'}\langle N_{\epsilon'}\rangle(0)\sum_{s=1}^{\infty}e^{-\lambda A^{2}t}\frac{(\lambda A^{2}t)^{s}}{s!\sqrt{s}}
\int_{-\infty}^{\infty}2zdzf(z^{2})e^{-\frac{2}{sk_{B}T_{CSL}}(z-\sqrt{\epsilon'})^{2}}
\end{eqnarray}
\noindent where $\overline f(t)\equiv\int_{0}^{\infty} d\epsilon f(\epsilon)\langle N_{\epsilon}\rangle(t)$. 

If $f=1$, so $\overline f\equiv\langle N\rangle(t)$ is the total number of particles in all states, and $\langle N\rangle(0)\equiv N$, one obtains:
\begin{eqnarray}\label{CL1}
\langle N\rangle(t)&=&Ne^{-\lambda A^{2}t}\nonumber\\
&&+\frac{1}{\sqrt{\pi k_{B}T_{CSL}/2}}\Big[1-e^{-\lambda A^{2}t}\Big]\int_{0}^{\infty}d\sqrt{\epsilon'}\langle N_{\epsilon'}\rangle(0)
2\sqrt{\epsilon'}\sqrt{\pi k_{B}T_{CSL}/2}=N
\end{eqnarray}
\noindent  so particle number is conserved. 

If $f=\epsilon=z^{2}$, so  $\overline f(t)=\overline E(t)$, one obtains:
\begin{eqnarray}\label{CL2}
\overline E(t)&=& \overline E(0)e^{-\lambda A^{2}t}\nonumber\\
&+&\int_{0}^{\infty}d\sqrt{\epsilon'}\langle N_{\epsilon'}\rangle(0)\sum_{s=1}^{\infty}e^{-\lambda A^{2}t}\frac{(\lambda A^{2}t)^{s}}{s!}
2\Big[3\sqrt{\epsilon'}\frac{sk_{B}T_{CSL}}{4} +\epsilon'^{3/2}\Big]\nonumber\\
&&\negmedspace\negmedspace\negmedspace\negmedspace\negmedspace\negmedspace\negmedspace\negmedspace\negmedspace\negmedspace
\negmedspace\negmedspace\negmedspace\negmedspace\negmedspace\negmedspace\negmedspace\negmedspace\negmedspace\negmedspace\negmedspace\negmedspace
=\overline E(0)e^{-\lambda A^{2}t}+\int_{0}^{\infty}d\epsilon'\langle N_{\epsilon'}\rangle(0)
\Big[\lambda A^{2}t\frac{3k_{B}T_{CSL}}{4} +\epsilon'(1-e^{-\lambda A^{2}t})\Big]=N\lambda A^{2}t\frac{3k_{B}T_{CSL}}{4}+\overline E(0),\nonumber\\
\end{eqnarray}
\noindent so the result is the linear energy increase Eq.(\ref{Rate2.5}).

\subsection{Initial BEC}

Consider an initial BEC of $N$ atoms and suppose an infinite box or harmonic trap. Then, the CSL excitation without collisions (and the attendant thermal equilibrium of the created cloud) is described by Eq.(\ref{C3c}). Here,  it is best to think of the BEC state as the
sum of two terms, one the initially populated state which decays, and the other, the lowest
member of the continuum of states. Using $\epsilon_{1}=\pi^{2}/2m\sigma^{2}=\alpha'^{2}k_{B}T_{CSL}/2$,  we obtain from Eq.(C3c):
\begin{subequations}
\begin{eqnarray}\label{CBEC1}
 \langle N_{\epsilon_{1}}\rangle(t)
&=& Ne^{-\lambda A^{2}t},\label{CBEC1a}\\
 \langle N_{\epsilon}\rangle(t)&=&\frac{1}{\sqrt{\pi k_{B}T_{CSL}/2}}\int_{0}^{\infty}d\sqrt{\epsilon'}N\delta(\epsilon'-\epsilon_{1})\sum_{s=1}^{\infty}e^{-\lambda A^{2}t}\frac{(\lambda A^{2}t)^{s}}{s!\sqrt{s}}
\Big[e^{-\frac{2}{sk_{B}T_{CSL}}(\sqrt{\epsilon}-\sqrt{\epsilon'})^{2}}-e^{-\frac{2}{sk_{B}T_{CSL}}(\sqrt{\epsilon}+\sqrt{\epsilon'})^{2}}\Big]\nonumber\\
&=&\frac{\alpha'}{2\sqrt{\pi} \epsilon_{1}}N\sum_{s=1}^{\infty}e^{-\lambda A^{2}t}\frac{(\lambda A^{2}t)^{s}}{s!\sqrt{s}}
\Big[e^{-\frac{\alpha'^{2}}{s\epsilon_{1}}(\sqrt{\epsilon}-\sqrt{\epsilon_{1}})^{2}}-e^{-\frac{\alpha'^{2}}{s\epsilon_{1}}(\sqrt{\epsilon}+\sqrt{\epsilon_{1}})^{2}}\Big].
\label{CBEC1b}\\
&\approx&\frac{2\alpha'^{3}\sqrt{\epsilon}}{\sqrt{\pi} \epsilon_{1}^{3/2}}N\sum_{s=1}^{\infty}e^{-\lambda A^{2}t}\frac{(\lambda A^{2}t)^{s}}{s!s^{3/2}}
e^{-\frac{\alpha'^{2}}{s\epsilon_{1}}\epsilon},\label{CBEC1c}
\end{eqnarray}
\end{subequations}
\noindent where, in obtaining Eq.(\ref{CBEC1c}), the approximation $\epsilon_{1}<<\epsilon$ is made.


\begin{thebibliography}{99}
\bibitem{Limit} This limit arises because free particles undergo random walk during collapse, and the resulting acceleration makes a charged particle radiate.  Fu\cite{Fu} calculated the rate of photon emission for an electron to be $R=\lambda' \pi^{-1}(e^{2}/4\pi\hbar c)(\hbar/m_{e}ca)^{2}(1/E_{keV}$)counts/keV-sec$=
1.02\times 10^{-20}\lambda/E_{keV}$ counts/keV-sec, where $e^{2}/4\pi\hbar c=1/137.0$, $\lambda'=\lambda(m_{e}/m_{p})^{2}$, $m_{e}$ is the electron mass and $m_{p}$ is the proton mass (see also\cite{AR}).  Fu applied this rate to "spontaneous" X-ray emission by the 4 valence electrons in Ge.  
 With 8.24$\times10^{24}$ atoms in a kg of Ge, the rate is $R= 2.87\times10^{10}\lambda/E_{keV}$ counts/keV-kg-day.  The inequality  $R<R_{expt}$ 
 provides the upper limit $\lambda<3.5\times 10^{-11}R_{expt}E_{keV}$. Improved data (drawn from Fig. 1 of the third reference in \cite{massp}) gives 
the experimental limit on spontaneous radiation appearing in a highly shielded 253g of Ge observed for 337 days as $R_{expt}=(40,20)$counts/.1keV-85.24kg-days, 
or (4.7,2.35)counts/keV-kg-day at $E=(5, 10)$keV respectively.  Since $R_{expt}E_{keV}=23.5$ for both data, the resulting limit is $\lambda<8\times 10^{-10}\approx 
10^{-9}$sec$^{-1}$.
\end{thebibliography}
\end{document}